\documentclass[journal]{IEEEtran}

\usepackage{gensymb}
\usepackage{fancyhdr}
\usepackage{amsmath}
\usepackage{amssymb}
\usepackage{graphicx}
\usepackage{booktabs}
\usepackage{amsfonts}
\usepackage{multirow}
\usepackage{algorithm}
\usepackage{algorithmic}
\usepackage{mathrsfs}
\usepackage{bm}
\usepackage{exscale}
\usepackage{relsize}
\usepackage{setspace}
\usepackage{color}
\usepackage{comment}
\usepackage{ulem}
\usepackage{cite}
\usepackage{subfigure}

\newtheorem{theorem}{\textbf{Theorem}}
\newtheorem{remark}{Remark}

\DeclareMathOperator*{\argmax}{arg\,max}

\ifCLASSINFOpdf
\else
\fi

\begin{document}

\title{Spatial-chirp Codebook-based Hierarchical Beam Training for Extremely Large-Scale Massive MIMO}

\author{
		Xu~Shi,~\IEEEmembership{Graduate~Student~Member,~IEEE},	Jintao~Wang,~\IEEEmembership{Senior~Member,~IEEE},
		Zhi~Sun,~\IEEEmembership{Senior~Member,~IEEE},
	     and Jian~Song,~\IEEEmembership{Fellow,~IEEE}
	  \thanks{
	  	Part of this work has been accepted by IEEE ICC 2023 \cite{shixu_ICC}.	
	  	This work was supported in part by Tsinghua University-China Mobile Research Institute Joint Innovation Center. (Corresponding author: Jintao Wang.)

	  	Xu Shi, Jintao Wang, Zhi Sun, and Jian Song are with the Department of Electronic Engineering, Tsinghua University, Beijing 100084, China and Beijing National Research Center for Information Science and Technology (BNRist). (e-mail: shi-x19@mails.tsinghua.edu.cn).
	  }
}


\maketitle

\pagestyle{empty}
\thispagestyle{empty}

\begin{abstract}
Extremely large-scale multiple-input multiple-output (XL-MIMO) promises to provide ultrahigh data rates in millimeter-wave (mmWave) and Terahertz (THz) spectrum. However, the spherical-wavefront wireless transmission caused by large aperture array presents huge challenges for channel state information (CSI) acquisition and beamforming. Two independent parameters (physical angles and transmission distance) should be simultaneously considered in XL-MIMO beamforming, which brings severe overhead consumption and beamforming degradation. To address this problem, we exploit the near-field channel characteristic and propose two low-overhead hierarchical beam training schemes for near-field XL-MIMO system. Firstly, we project near-field channel into spatial-angular domain and slope-intercept domain to capture detailed representations. Then we point out three critical criteria for XL-MIMO hierarchical beam training. Secondly, a novel spatial-chirp beam-aided codebook and corresponding hierarchical update policy are proposed. Thirdly, given the imperfect coverage and overlapping of spatial-chirp beams, we further design an enhanced hierarchical training codebook via manifold optimization and alternative minimization. Theoretical analyses and numerical simulations are also displayed to verify the superior performances on beamforming and training overhead.
\end{abstract}


\IEEEpeerreviewmaketitle

\section{Introduction} 

Massive Multiple-Input Multiple-Output (MIMO) is acknowledged as one of the key ingredients for the fifth generation (5G) of wireless communications, which has been standardized as 5G New Radio (NR) \cite{XLMIMO_1,XLMIMO_2,XLMIMO_3,XLMIMO_4}. As the communication frequency-band is further extended to millimeter-wave (mmWave) and Terahertz (THz) spectrum, extremely large-scale massive MIMO (XL-MIMO) with significant number of antennas is promising to provide much stronger beamforming gain and higher spectrum efficiency \cite{XLMIMO_3}. However, caused by the large aperture arrays and corresponding high frequency band in XL-MIMO, Rayleigh distance may appear up to several hundred meters, which means the base station (BS) will serve large near-field (i.e., Fresnel region) areas inside Rayleigh distance \cite{nearfield_1,minimum_distance05}. 

Specifically, spherical-wavefront assumption instead of conventional planar wavefront should be reconsidered in near-field scenario. The corresponding channel is characterized by two independent parameters, i.e., the angle-of-departure/arrival (AoD/AoA) and transmission distance \cite{nearfield_distance_2}. Under the specific near-field spherical-wavefront channel conditions, the transceiver, beamforming codebooks and CSI acquisition should be all redesigned for correct pair-matching of wireless channel. No matter whether analog or hybrid beamforming structures are adopted, conventional far-field beamforming codebook will suffer from deadly gain degradation and costly time complexity \cite{CE_2_polardomain}. Moreover, the extremely large antennas and distance-sensitive channel steering response will herein bring about severe training overhead. How to design the near-field beamforming codebook for XL-MIMO with low training overhead is not addressed to our best knowledge. 

\subsection{Related works}
The previous studies for near-field spherical-wavefront wireless electromagnetic radio transmission can be broadly separated into two categories: (a) One is sensing and localization of near-field sources \cite{sensing_1_subspace,sensing_2_2DMUSIC,sensing_3_ML,sensing_4_harmonics}. Many maximum likelihood methods \cite{sensing_3_ML} and subspace-based methods were widely studied such as modified 2-dim multiple signal classification (2D-MUSIC) \cite{sensing_2_2DMUSIC} and the spherical harmonics domain method \cite{sensing_4_harmonics}. Most methods in near-field localization and sensing involve large covariance matrix, higher-order statistics and multidimensional searching. Therefore, those methods are unsuitable for XL-MIMO beam training due to the huge computational complexity and overhead consumption. (b) The other related research direction is near-field sparse channel estimation via compressive sensing (CS) \cite{CE_1_subarray,CE_2_polardomain,CE_3}. \cite{CE_1_subarray} uniformly divides the large aperture antenna array into several subarrays, and proposes a refined orthogonal matching pursuit (OMP)-based subarray/scatterer-wise channel estimation method. Furthermore, \cite{CE_2_polardomain} exploits the polar-domain sparsity of near-filed XL-MIMO channel and designs a polar-domain simultaneous OMP algorithm for explicit CSI acquisition. However in CS-based estimation, the received signal-to-noise ratio (SNR) at user equipment (UE) side is usually quite low since the beamforming matrix needs random generation for restricted isometry property \cite{CS_RIP_1}, which may cause CSI acquisition degradation. In another word, the implicit CSI acquisition, i.e., near-field beam training, is also significant and seems more realistic and reliable for XL-MIMO. 

Conventional far-field hierarchical beamspace training has been widely developed in both academic research and standardization progress. 
A predefined hierarchical codebook including several layers of codebooks is typically employed, where the spatial region covered by the codeword at an upper layer is split into several smaller spatial regions covered by codewords at lower layers.  For example,
in IEEE 802.11ad/802.15.3c \cite{standardIEEE} sector level sweep (SLS), beam-refinement protocol (BRP) and beam tracking were definitely settled. Besides, from an academic aspect, \cite{FFhierarchical_1_JOINT} utilized hierarchical weighted summation of sub-arrays and proposed a joint sub-array and de-activation (JOINT) hierarchical codebook. Enhanced JOINT (EJOINT) method was further proposed in \cite{FFhierarchical_2_EJOINT} to avoid antenna de-activation with improved training success rate. Furthermore, Riemannian optimization-based method \cite{FFhierarchical_3_manifold} and successive closed-form (SCF) algorithm \cite{FFhierarchical_4_SCF} were also adopted for efficient beam coverage. 
Those works mainly focus on the planar-wavefront beam pattern design but neglect the hierarchical update policy optimization, where binary update/searching is widely adopted by default due to the only $1$-dim angular training in far field. However, in the spherical-wavefront XL-MIMO scenario, not only the angular parameter but additional distance-dependent factor should be searched for accurate beamfocusing, which leads to excessive training overhead and computational complexity. Correspondingly, the spherical-wavefront hierarchical training, especially the $2$-dim layer-by-layer update policy and unified near-field beam pattern, should be carefully redesigned to match the joint distance-angular-related searching in near-field scenario.

Up to now, there exists limited research for near-field hierarchical beam searching. Distance-based beam training for XL-MIMO is proposed in \cite{weixiuhong}, where the codewords are determined by a pair of uniformly sampled points in realistic space coordinate system. Hierarchical layers are controlled via different lengths of sampling spacing. The distance-based scheme is intuitive but not optimal unfortunately. 
Accordingly, nearby regions of BS may suffer from insufficient resolution due to its distance sensitivity while distance-insensitive far-field regions will be deployed with redundant codewords. In another word, the distance-based codebook causes unfair codeword allocation and severe searching precision degradation. Furthermore, the codebook size will sharply boost as transmission distance raises, which is unacceptable for realistic communication.
Another fast training scheme is proposed in \cite{fast_training} which first searches angular domain and then traverses overall distance rings in the second phase. The inter-beam interference and relevance are not thoroughly considered in \cite{fast_training}, which may induce redundant training overhead. In fact, inter-beam relevance among several distances promotes hierarchical training design by the flexible wide area coverage, which is the basic motivation of this work, with enhanced training success rate and further reduced overhead fortunately.

\subsection{Contributions}
As analyzed above, the main challenges of near-field hierarchical training are the excessive overhead and lack of self-contained hierarchical update policy. The former results from the additional distance-dependent parameter besides angular value, and the latter roots in the ineffectiveness of binary searching in the near-field XL-MIMO training.
Specifically, this article tries to solve the following three critical puzzles in near-field XL-MIMO hierarchical beam training, including: (1) how to find the unified near-field beam pattern and obtain elementary codebook; (2) how to generate multi-resolution codebooks and design the self-contained hierarchical update policy for overhead reduction; (3) how to give the theoretical analysis of multiple resolutions and overhead consumption.

The main contributions of this paper are summarized as follows:
\begin{itemize}
	\item We provide \emph{joint spatial-angular} and \emph{slope-intercept} representations for near-field spatial non-stationary channel. The near-field steering vector (spatial domain) contains the same mathematical expression, i.e., linear and quadratic phase terms, compared with the linear frequency modulation signal (time domain). Inspired by Joint Time-Frequency Analysis (JTFA) for non-stationary analysis, we project the near-field beam steering vector into $2$-dim spatial-angular plane and obtain its spatial non-stationary characteristic. Just like that far-field beam is mapped into one point in $1$-dim beamspace, we herein project each near-field steering vector into one point at $2$-dim slope-intercept domain and obtain the elementary codebook for XL-MIMO beam training. 
	
	\item We further propose \emph{three critical criteria} for XL-MIMO hierarchical beam training, which includes the requirements of multi-resolution codebook coverage, inter-codeword interference and generation complexity.
	
	\item Motivated by the characteristic of chirp signal, we propose a novel spatial-chirp-based hierarchical beam training scheme for near-field XL-MIMO.
	Benefitted from the spatial-nonstationary characteristic, each spatial-chirp beam plays a dual role inside beam training: It is both one narrow beam in its present distance ring, and also wide beam for other distance rings, which means it can cover one wide-area range with dynamic area related to transmission distance, and this helps reduce training overhead from exhaustive searching. Furthermore, the hierarchical update policy can be redesigned corresponding to the specific distance-related dominant area for each near-field beam.
	
	\item Given the gap between ideal beam pattern and realistic spatial-chirp beam, we further propose an enhanced spatial-chirp-based hierarchical training scheme. Alternative minimization and Riemannian manifold optimization are adopted for gap reduction of near-field beam pattern, where the overlapping among spatial-chirp beams is eliminated and training success rate is significantly improved. 
	
	\item Theoretical analyses of hierarchical layer partition, beam training overhead and computational complexity are derived in this paper. Numerical results are provided to verify the effectiveness of spatial-chirp-based hierarchical training for near-field XL-MIMO. Training overhead can be reduced by over $99\%$ with slight beamforming performance degradation.
\end{itemize}

\subsection{Organization and Notations}
The rest of this paper is organized as follows. In Section \uppercase\expandafter{\romannumeral2}, we introduce the XL-MIMO system and analyze near-field spherical-wavefront channel. In Section \uppercase\expandafter{\romannumeral3}, we give joint spatial-angular analysis of near-field channel and propose the slope-intercept domain representation of spatial non-stationary beams. The spatial-chirp-based hierarchical beam training scheme is proposed in Section \uppercase\expandafter{\romannumeral4}, while we further design an enhanced scheme for overlapping elimination in Section \uppercase\expandafter{\romannumeral5}. Finally, we show the simulation results in Section \uppercase\expandafter{\romannumeral6} and conclude this paper in Section \uppercase\expandafter{\romannumeral7}.

\textit{Notations}: Lower-case and upper-case boldface letters $\mathbf{a}$ and $\mathbf{A}$ denote a vector and a matrix respectively. $\otimes$ and $\circ$ denote the Kronecker product and Hadamard product, respectively. $\mathbf{I}_M$ is identity matrix with size $M\times M$. The operator $\text{diag}(\mathbf{a})$ is a square diagonal matrix with entries of $\mathbf{a}$ on its diagonal. The operators $\text{vec}(\cdot)$ and $\text{mat}(\cdot)$ denote the vectorizing of a matrix and matrix-reshape of a tensor, respectively. $||\bm\Delta||_F$ and $||\bm\Delta||_2$ denote Frobenius norm and $l_2$ norm for matrix $\bm\Delta$. $\mathcal{CN}(\bm{\mu},\mathbf{\Sigma})$ stands for complex Gaussian distribution with mean $\bm{\mu}$ and variance $\mathbf{\Sigma}$. $\text{rect}(2a)$ denotes a rectangular window with supporting interval $[-a,a]$ and $\angle\mathbf{a}$ denotes the element-wise phase extraction of vector $\mathbf{a}$.

\section{System Model}
\subsection{System and Problem Description}

\begin{figure}[!t]
\centering
\includegraphics[width=0.8\linewidth]{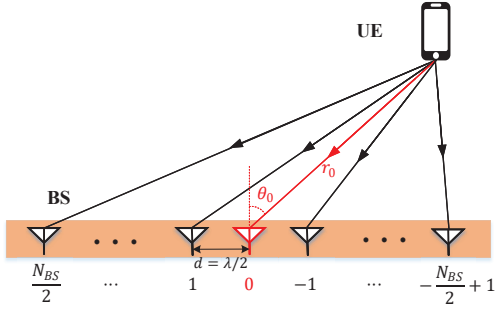}
\caption{Block diagram of near-field XL-MIMO system model.}
\label{system_model}
\end{figure}

We consider a narrow-band downlink XL-MIMO system. Base station (BS) is configured with $N_\text{BS}$-element uniform linear array (ULA) with antenna indices as $n\in \{-\frac{N_\text{BS}}{2}+1,\dots,\frac{N_\text{BS}}{2}\}$. We consider one RF chain at BS side (i.e., analog beamforming structure) in this paper\footnote{Hybrid precoding structure can be similarly extended with more flexible beam pattern design.}. Without loss of generality, we only consider one single-antenna user equipment (UE) as shown in Fig. \ref{system_model}. The distance and corresponding AoD between UE and BS array center (i.e., antenna element $n=0$) are marked as $r_0$ and $\theta_0$ \footnote{In the following content, we only pay attention on AoD sine value and thus replace $\sin \theta_0$ with $\theta_0\in [-1,1]$ in array manifold for convenience.}, respectively. The carrier frequency is set as $f_c$. Denote $d=\lambda/2$ as the fixed antenna spacing and $\lambda=c/f_c$ is the wavelength of electromagnetic waves. The overall ULA size is marked as $D=d(N_\text{BS}-1)\approx dN_\text{BS}$. Therefore, the received noisy signal at UE can be formulated as 
\begin{equation}
y_t = \mathbf{h}^T\mathbf{f}_t s_t+n_t,
\end{equation}
where $\mathbf{f}_t\in \mathbb{C}^{N_\text{BS}\times 1}$ represents phase shifter (PS)-aided beamforming vector in the $t$-th timeslot and $n_t\in \mathcal{CN}(0,\sigma_N^2)$ is additive white Gaussian noise (AWGN) with power $\sigma_N^2$. $\mathbf{h}$ denotes the mmWave wireless channel and is written as 
\begin{equation}
\mathbf{h} = \beta_\text{LoS}\mathbf{a}_\text{LoS}+ \sum_{l=1}^{N_\text{NLoS}}\beta_l \mathbf{a}_l,
\end{equation}
$\beta$ here is complex path loss and $\mathbf{a}$ is the corresponding array steering vector with each entry $a_n=e^{-j2\pi r_n/\lambda}$. For LoS path, $r_n$ denotes the distance between the $n$-th BS antenna and UE, while for NLoS path, $r_n$ represents the distance between scatterer and $n$-th BS antenna. 

The objective of beam training is to select the optimal codeword $\mathbf{f}_\text{opt}$ from finite codebook $\mathcal{F}$ to maximize the system spectral efficiency or beamforming gain. And the problem can be formulated as follows:
\begin{equation}
\arraycolsep=1.0pt\def\arraystretch{1.0}
\begin{array}{rlc}
\displaystyle \max_\mathbf{f} && |\mathbf{h}^T\mathbf{f}|^2\\
\text{s.t.} &&\ \  \mathbf{f}\in \mathcal{F}\ \  \text{and}\ \  |\mathbf{f}_n|=1,\forall n
\end{array}.
\end{equation}
Generally, the LoS path gain $\beta_\text{LoS}$ is much larger than NLoS gain $\beta_l$ especially when carrier frequency is high enough such as mmWave and THz scenarios, and thus in beam training we mainly focus on the strongest LoS beam searching.
\subsection{XL-MIMO Channel Approximation}

The main difference between XL-MIMO channel and conventional MIMO lies in the wavefront pattern of beam steering vector $\mathbf{a}_\text{LoS}$ as follows.
The conventional far field is dominated by radiated fields and the direction of propagation is assumed as with plane waves. Conversely, in the near-field (Fresnel) region, the shape of radiation pattern varies with distance, i.e., spherical-wavefront propagation. The boundary value between far-field and Fresnel regions can be defined as Rayleigh distance $r_\text{R}=\frac{2D^2}{\lambda}$.
Since the carrier frequency $f_c$ and the number of antenna turn extremely large, the Rayleigh length further increases and even gets larger than realistic supporting communication distance $r_0$, which means that the far-field plane-wave assumption couldn't hold anymore. The transmission distance for $n$-th antenna element should be exactly calculated via cosine rule as:
\begin{equation}
r_n = \sqrt{r_0^2+(nd)^2+2 r_0 nd\theta_0}\overset{(a)}{\approx} r_0+\theta_0 \cdot nd+\frac{1-\theta_0^2}{2r_0}\cdot (nd)^2,
\end{equation} 
where $(a)$ is approximated via Taylor Expansion, which has been widely adopted in previous near-field spherical-wave propagation model \cite{nearfield_1}. Notice that the first term is common to all antenna elements and is neglected here, the second term corresponds to conventional plane-wave array and the third term here is an additional component in spherical-wave XL-MIMO. Therefore the XL-MIMO array steering response $a_n=e^{-j2\pi r_n/\lambda}$ is approximate to \footnote{ We only consider the phase change of steering response and neglect the amplitude change in this article. The further detailed near-field model and applicable region analysis can be found in \cite{sunshu}. }
\begin{equation}
	\arraycolsep=1.0pt\def\arraystretch{1.5}
\begin{array}{ccl}
\hat{a}^\text{n-f}_n &=&\displaystyle \text{exp}\left\{-j\pi\left (\theta_0 n+\frac{\lambda (1-\theta_0^2)}{4r_0}n^2\right)\right\}\\
&=&\displaystyle \text{exp}\left\{-j\pi\left (\theta_0 +\frac{\lambda (1-\theta_0^2)}{4r_0}n\right)\cdot n\right\}
\end{array}
\label{app_near_field_beam}
\end{equation}

Correspondingly the near-field XL-MIMO beam training problem can be rewritten as 
\begin{equation}
\arraycolsep=1.0pt\def\arraystretch{1.0}
\begin{array}{rlc}
\displaystyle \max_\mathbf{f} && |\hat{\bm a}_\text{LoS}^{\text{n-f},T}\mathbf{f}|^2\\
\text{s.t.} &&\ \  \mathbf{f}\in \mathcal{F}\ \  \text{and}\ \  |\mathbf{f}_n|=1,\forall n
\end{array}.
\label{problem_formulation}
\end{equation}
In a word, we need to pick the best beam from the predefined codebook by measuring the scalar powers corresponding to each codeword. Although the problem itself shares the same style with conventional far-field beam training, entirely new challenges exist in the spherical-wavefront beam training due to the more complicated channel characteristic. Not only the AoD value but the distance-dependent quadratic term in (\ref{app_near_field_beam}) should be considered in the codebook design $\mathcal{F}$ to match the near-field channel $\hat{\bm a}_\text{LoS}^\text{n-f}$, which causes severe codebook-size multiplication.
Although distance-based codebooks \cite{weixiuhong} partially resolves the excessive overhead problem, the codebook coverage and the layer-by-layer hierarchical update policy are still imperfect and unfair for different user locations. Due to the uniform quantization of physical spatial domain in \cite{weixiuhong}, nearby regions of BS may suffer from insufficient resolution due to its distance sensitivity while distance-insensitive far-field regions will be deployed with redundant codewords. The ingenious hierarchical beam training in near-field XL-MIMO, with joint consideration of \emph{low overhead consumption}, \emph{low computational complexity} and \emph{perfect beam coverage}, has not been perfectly addressed to our best knowledge.

\section{Elementary Codebook Design and $k-b$ Domain Representation}

\begin{figure}[!t]
\centering
\includegraphics[width=0.7\linewidth]{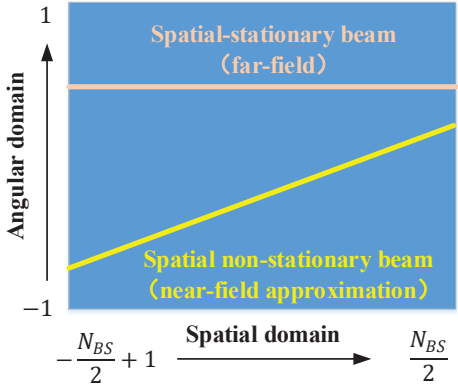}
\caption{Comparison of joint spatial-angular analysis for spatial stationary/ non-stationary beam vectors.}
\label{non_stationary_beam}
\end{figure}

Inspired by linear frequency modulation (LFM) signal $e^{-j\pi (f_0+kt)t}$ (also named as chirp signal) in continuous wave radar, we can easily observe that the approximated near-field  beam (\ref{app_near_field_beam}) has the same structure. The instantaneous AoD at each antenna element $\theta_n = \theta_0 +\frac{\lambda (1-\theta_0^2)}{4r_0}n$  increases linearly with slope $k$ and intercept $b$ as:
\begin{equation}
k=\frac{\lambda (1-\theta_0^2)}{4r_0}, \ \ \ b=\theta_0.
\label{k_b_formula}
\end{equation}
It is worth noting that the phase formulation $\theta_n=\theta_0+\frac{\lambda (1-\theta_0^2)}{4r_0}n$ is a sufficient description for near-field steering vector representation just like LFM signal.
However, notice that the near-field beam searching is quite different and much more challenging compared with LFM signal detection, this is because the spatial-sampling  rate ($1/d$) and observation number (controlled by pilot length $T$) are both limited in  beam searching. Thus conventional LFM detection schemes such as short-time Fourier transform (STFT) and ambiguity function cannot be adopted directly in XL-MIMO beam training. Compared with far-field antenna steering response $a^\text{f-f}_n=e^{-j\pi\theta_0n}$ with $k=0$, the near-field beam seems a classical spatial non-stationary signal. For more readability and easier analysis for channel characteristic, like Joint Time-Frequency Analysis (JTFA), we project the near/far-field beam vectors $\mathbf{\hat{a}}^\text{n-f}=[\hat{a}^\text{n-f}_n]$ and $\mathbf{a}^\text{f-f}=[\hat{a}^\text{f-f}_n]$ into spatial-angular plane as shown in Fig. \ref{non_stationary_beam}. Inside the spatial-angular representation, we give a visual description for the phase variation, where horizontal and vertical axes denote the antenna indices $n$ and instantaneous AoD $\theta_n$ respectively. For the far-field beam, all antenna elements contain the same direction to user location, i.e., the horizontal line in Fig. \ref{non_stationary_beam}. Conversely in the near-field case, all antenna elements own individual directions pointing to user location, which varies linearly with antenna indices, and we name it as spatial non-stationary beam.

As for conventional far-field beam, since the overall beams' slope is fixed as $k=0$, the searching of all AoD values (intercepts) is enough for the best beamforming direction selection. Furthermore, the conventional hierarchical searching is also intuitive: In the top hierarchical layer, wide beams are equipped to coarsely search for a larger angular interval, and then in the next hierarchical layers the large sector is split into several small fractions for more accurate beam training. In a word, the conventional  codebooks can be easily generated by fully covering the angular domain $-1\leq \theta_0\leq 1$ with minimum beam interference. 

On the contrary, in XL-MIMO model, we have to simultaneously estimate both the slope $k$ and intercept $b$ to determine the optimal spatial-chirp beam (\ref{app_near_field_beam}), which causes more pilot consumption and tremendous challenge in hierarchical codebook design. Different from previous studies that mainly focus on direction-distance-based codebook design, we herein decouple the two parameters and consider the direct quantization of $k$ and $b$. Notice that the group $(k,b)$ is equivalent to $(\theta_0,r_0)$ due to its injective property, but $(k,b)$ is more general and low-complexity compared with $(\theta_0,r_0)$ because of the decoupled relationship in spatial-chirp beams.

\begin{figure*}[!t]
\centering
\includegraphics[width=0.7\linewidth]{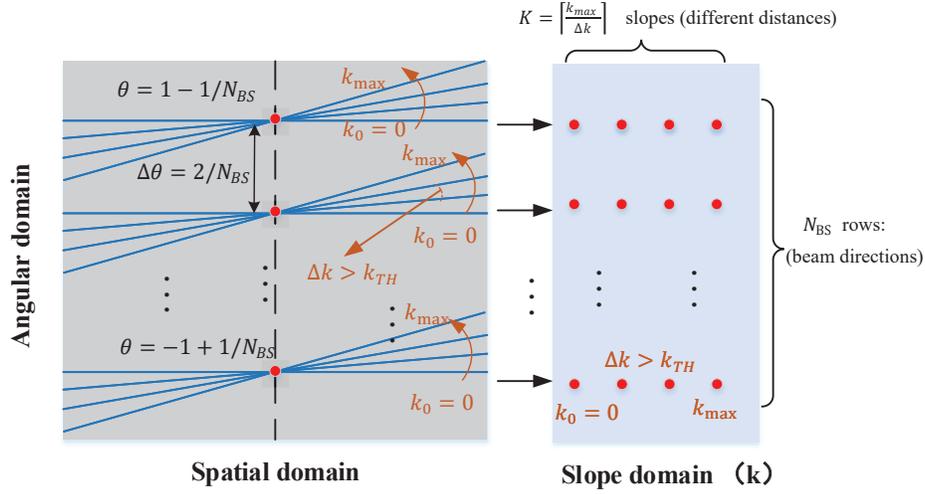}
\caption{Elementary codebook for near-field XL-MIMO in spatial-angular domain and slope-intercept ($k$-$b$) domain representation.}
\label{k_b_representation}
\end{figure*}

Then a trivial and elementary codebook can be generated by uniform quantization of $k$ and $b$ as shown in Fig. \ref{k_b_representation}. First the intercept interval $[-1,1]$ are uniformly quantized to $N_\text{BS}$ groups. Inside each quantized intercept $b_q$, several quantized slopes are independently modulated to form different spatial-chirp beams. The slope interval is marked as $[k_\text{min},k_\text{max}]$ where $k_\text{min}=0$ corresponds to maximum transmission distance $r\rightarrow +\infty$ and $k_\text{max}=\frac{\lambda}{4r_\text{min}}$ corresponds to the minimum BS serve distance $r_\text{min}$. The slope quantization spacing is defined as $\Delta k<k_\text{TH}$ and the threshold $k_\text{TH}$ will be given in the following sections. Considering beams $\mathbf{\hat{a}}_1,\mathbf{\hat{a}}_2$ with common slope $k$ and respective intercept $\theta_1$ and $\theta_1+p\frac{2}{N_\text{BS}} \ (p\in \mathbb{Z})$, the beam coherence is strictly orthogonal as 
\begin{equation}
    g_\text{b}=|\mathbf{\hat{a}}_1^H\mathbf{\hat{a}}_2|=\sum_{n=1-N_\text{BS}/2}^{N_\text{BS}/2}e^{-j\pi p\frac{2}{N_\text{BS}}n}=N_\text{BS}\cdot \delta(p).
\end{equation}
Considering beams $\mathbf{\hat{a}}_3,\mathbf{\hat{a}}_4$ with common intercept $b$ and respective intercept $p \Delta k$ and $q\Delta k \ (p,q\in \mathbb{Z})$, the beam coherence is derived as 
\begin{equation}
g_\text{k}=|\mathbf{\hat{a}}_3^H\mathbf{\hat{a}}_4|=\sum_{n=1-N_\text{BS}/2}^{N_\text{BS}/2}e^{-j\pi (q-p)\Delta k n^2}.
\end{equation}
which is a discrete Fresnel summation function with variable $(q-p)\Delta k$. Although the two beams here are not exactly orthogonal, we can control the coherence (interference) value via parameter design of $\Delta k$ according to \cite{CE_2_polardomain}.

Next, we transform the spatial-angular domain into slope-intercept ($k$-$b $) plane as shown in Fig. \ref{k_b_representation}.  It is worth noting that the two representations here are equivalent but with different  manifestations. The spatial-angular representation is to curve the instantaneous AoDs $\theta_n = \theta_0 +\frac{\lambda (1-\theta_0^2)}{4r_0}n$ via variables $(\theta_n,n)$, while the $k$-$b$ domain is to describe the phase variation via parameters $(k,b)=(\frac{\lambda (1-\theta_0^2)}{4r_0},\theta_0)$. The latter seems more significant since beam coherence with different parameters can be observed easily in the $k$-$b$ domain representation. The detailed transform procedure is as follows. Each spatial-chirp beam (line in Fig. \ref{k_b_representation}) is projected to one point with coordinate $(k,b)$ and the whole codebook is just the uniform quantization of the 2-dim plane $[-1,1]\times [k_\text{min},k_\text{max}]$, defined as $\mathcal{F}_\text{ele}=\{\bm w_{k,b}\}$. A straightforward method to solve (\ref{problem_formulation}) in XL-MIMO is the exhaustive searching of $\mathcal{F}_\text{ele}$, but we can obviously observe that the codebook size is quite huge and unacceptable for realistic traversal searching. For example, when $N_\text{BS}=1024$, $f_c= 100 \text{GHz}$ and $r_\text{min}=10\text{m}$, the corresponding quantized number at slope axis and codebook size are $16$ and $1024\times 16=16384$, respectively. Therefore the hierarchical spatial-chirp beam training and corresponding codebook design scheme are necessary for low pilot consumption. Notice that the slope-intercept plane is the basis and kernel of the following content. Hierarchical codebook design, enhancement and the extension for other specific scenarios are all based on the $k-b$ domain. Motivated by the idea in \cite{FFhierarchical_2_EJOINT}, we herein provide \emph{three basic criteria} for XL-MIMO hierarchical beam training as follows.

\emph{\textbf{Criterion 1 (intra-layer):}}
 The $k$-$b$ domain region union supported by overall codewords in each hierarchical layer, should cover the whole $k$-$b$ domain $[k_\text{min},k_\text{max}]\times [-1,1]$, i.e.,
\begin{equation}
\cup_{n=1}^{N^{(l)}}\left[\ \mathcal{R}( \bm w_n^{(l)})\  \right] = [k_\text{min},k_\text{max}]\times [-1,1], l=1,\dots,L. 
\end{equation} 
where $\bm w_n^{(l)}$ represents the $n$-th codeword in the $l$-th hierarchical layer. Assume the whole codebook at layer $l$ is $\mathcal{F}^{(l)}=\{\bm w^{(l)}_n\}$. Then $\mathcal{R}(\bm w^{(l)}_n)$ denotes the  dominant region of the $n$-th beam inside the $l$-th layer, and can be mathematically expressed as 
\begin{equation}
\mathcal{R}(\bm w_n^{(l)})=\left\{(k,b) \bigg|  \bm w_n^{(l)}=\argmax_{\bm w\in \mathcal{F}^{(l)}} |\bm w^H\mathbf{a}^\text{n-f}_{(k,b)}|\right\}
\end{equation}

\emph{\textbf{Criterion 2  (inter-layer):}} The coverage region of any codeword $\bm w^{(l)}_n$, should be completely covered by the region union of several codewords in the next layer, i.e.,
\begin{equation}
	\mathcal{R}(\bm w_n^{(l)})\subseteq  \cup_{m\in \mathcal{I}_n^{(l)}}\left[ \mathcal{R}(\bm w^{(l+1)}_m)\right]
\end{equation}
where $\bm w_n^{(l)}$ denotes the $n$-th codeword in the $l$-th layer, and $\mathcal{I}_n^{(l)}$ represents the index set containing indices of the codewords in the $(l+1)$-th layer, which are split by $\bm w_n^{(l)}$ in the upper layer.

\emph{\textbf{Criterion 3 (scalability):}}  Inside each layer, the rest codewords can be yielded from one codeword via beam rotation, which also means the dominant $k$-$b$ domain regions for all codewords should contain the same size and geometric shape, i.e.,
\begin{equation}
	\arraycolsep=1.0pt\def\arraystretch{1.5}
\begin{array}{ccl}
\bm w^{(l)}_n &=&\displaystyle\bm w^{(l)}_1\circ \mathbf{a}^\text{n-f}_{(p_n\Delta k,q_n\Delta b)},\\
\mathcal{R}(\bm w^{(l)}_n)&=&\displaystyle \mathcal{R}(\bm w^{(l)}_1)+(p_n \Delta k, q_n \Delta b)
\end{array}
\end{equation}


To summarize, the above criteria aim to fully support and cover the overall $k-b$ domain $[-1,1]\times [k_\text{min},k_\text{max}]$ with interference as small as possible. For the slope-intercept ($k$-$b$) domain, the following property can be easily yielded, which is helpful for the following derivations:
\begin{theorem}
Given any steering vector $\mathbf{v}\in \mathbb{C}^{N_\text{BS}\times 1}$ with unit-modulus entries ($|v_n|=1$), for $k_0$-th column's overall $N_\text{BS}$ normalized codewords ($\bm w_{k,b}$ with all uniformly quantized $b\in [-1,1]$ and fixed $k=k_0$, $\|\bm w_{k,b}\|^2_F=1$), the summation of squared coherence is constant and equal to
\begin{equation}
\sum_{b}|\bm w_{k_0,b}^H\mathbf{v}|^2=N_\text{BS}, \ \forall k_0\in [k_\text{min},k_\text{max}]
\end{equation}
\label{Theorem1}
\end{theorem}
\begin{IEEEproof}
For any column of $k-b$ domain with $k=k_0$, the normalized codewords insides are formulated as 
\begin{equation}
    \bm w_{k_0,b_q}=\left[\frac{1}{\sqrt{N_\text{BS}}} \cdot e^{-j\pi (k_0 n^2 + b_q n)}\right],
\end{equation}
and the squared coherence summation is
\begin{equation}
	\arraycolsep=1.0pt\def\arraystretch{1.5}
\begin{array}{ccl}
 \sum_{b_q}|\bm w_{k_0,b_q}^H\bm v|^2&=& \displaystyle \sum_{b_q}\left|\sum_n \left(\frac{1}{\sqrt{N_\text{BS}}}e^{-j\pi k_0 n^2 } e^{-j\pi b_q n} v_n\right) \right|^2\\
 &=&\displaystyle \sum_{b_q}\left|\frac{1}{\sqrt{N_\text{BS}}} \sum_n v_n^\prime e^{-j\pi b_q n}  \right|^2
 \end{array}
    \label{parseval_equ}
\end{equation}
where $\bm v^\prime=[v_n e^{-j\pi k_0 n^2 }]$ is auxiliary vector. Notice the final result of (\ref{parseval_equ}) is just the overall power of $\bm v^\prime$ in frequency (angular) domain. According to Parseval's theorem \cite{WC_book}, we can get that the power in frequency (angular) domain is equal to the power in time (spatial) domain. Since $\bm v^\prime$ is still with all entries unit-modulus ($|v^\prime_n|=1$), the time (spatial) domain power is $\bm v^{\prime\ H}\bm v^\prime=N_\text{BS}$ and thus we finish the proof of Theorem \ref{Theorem1}.

Besides we can also obtain the same result via fractional Fourier Transform (FrFT) with different rotation angles.
\end{IEEEproof}

\begin{remark}
From Theorem \ref{Theorem1} we get that: Beamforming codewords contain the same power $N_\text{BS}$ no matter which column of $k-b$ domain we project it into. This also means that, we cannot design such a simple codeword that only scans for a fraction in slope $k$ axis. At least, it is quite difficult to find a beamforming vector that only supports a fractional square $[k_1,k_2]\times [b_1,b_2]$ (with the rest domain's coherence all zero) inside the whole domain $[k_\text{min},k_\text{max}]\times [-1,1]$. In another word, even if such codewords are designed, the corresponding inter-codeword interference and beam training overhead may further degenerate since the rest $k$-$b$ region's power is not fully considered or exploited.
\end{remark}

\section{Spatial-chirp-based Hierarchical Beam Training}
In this section, we propose a spatial-chirp-based hierarchical beam training scheme. Characteristics of spatial-chirp beam are fully analyzed and exploited to reduce redundant training overhead. Corresponding hierarchical update policy is also provided here.
Fortunately, the final training error probability, overhead consumption and average beamforming gain all get brilliant improvements through our proposed spatial-chirp-based hierarchical training scheme.

\subsection{Spatial-Chirp Beam Pattern Analysis in $k-b$ Domain}

\begin{figure}[!t]
	\begin{center}
		\subfigure[Spatial-domain beam pattern and angular-domain beam pattern of near-field LoS channel.]{
			\includegraphics[width=0.8\linewidth]{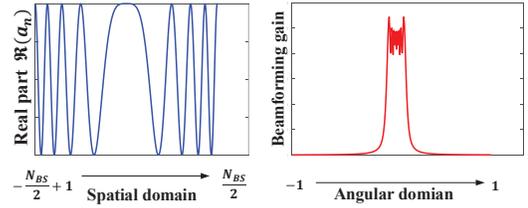}
			\label{each_spatial_angular_chirp_beam_pattern}
		}
		\subfigure[The ideal and realistic beam pattern of spatial-chirp beam in $k$-$b$ domain representation.]{
			\includegraphics[width=0.8\linewidth]{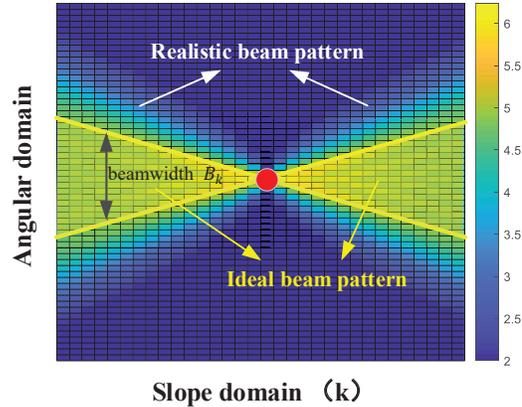}
			\label{spatial_chirp_beam_pattern}
		}
		\caption{The near-field LoS channel representation in spatial domain, angular domain and $k$-$b$ domain.}
	\end{center}
\end{figure}

We herein give an analysis for spatial-chirp beam in $k$-$b$ domain first, which is significant for both far-field and near-field hierarchical beam training. According to Theorem \ref{Theorem1}, spatial-chirp beams contain the same power along all columns' coherence, but the detailed coherence distribution inside $k-b$ domain is not captured yet. Taking one spatial-chirp beam with parameters $k=k_0$ and $b=b_0$ as example, the mathematical expression $\mathbf{f}_\text{eg}(k_0,b_0)$ is written in (\ref{chirp_analysis}), while the corresponding spatial-domain curve (real component) and angular-domain (with $k=0$) curve are shown in Fig. \ref{each_spatial_angular_chirp_beam_pattern}. 
\begin{equation}
	\arraycolsep=1.0pt\def\arraystretch{1.5}
\begin{array}{ccl}
&\mathbf{f}_\text{eg}(k_0,b_0) =&\displaystyle \left[e^{-j\pi\left(k_0(-\frac{N_\text{BS}}{2}+1)^2 + b_0 (-\frac{N_\text{BS}}{2}+1)\right)},\dots\right.
\left. ,\right.\\
&&\displaystyle\ \ \ \ \ \ \ \ \left. 1, \dots,e^{-j\pi\left(k_0(\frac{N_\text{BS}}{2})^2 + b_0 \frac{N_\text{BS}}{2}\right)}\right]^T
\end{array}
\label{chirp_analysis}
\end{equation}
One key indicator of a spatial-chirp beam is the bandwidth. Firstly, we extend the beam width from conventional angular (beamspace) domain to the 2-dim $k-b$ domain. At each column with $k=k_1$, the spatial-chirp beam will appear as individual beam pattern and beam width. We can define the beam width at each column $k=k_1$ following the $3\text{dB}$ rule, marked as $B_{k_1}$. The $3\text{dB}$ bandwidth here is known as the half-power bandwidth, at which point the output power has dropped to half of its peak value. Since the precise formulation of the bandwidth is complicated and counter-intuitive for hierarchical training design, we use the principle of stationary phase instead for a simplified approximate evaluation as follows:
\begin{equation}
B_{k_1} \approx |k_0-k_1|N_\text{BS}
\label{ideal_pattern_width}
\end{equation}

When $k_1=k_0$, the scanning interval will degrade to one point, i.e, $(k_0,b_0)$ in slope-intercept plane with power $N_\text{BS}$. Notice that each beam contains a bandwidth and thus the beam point $(k_0,b_0)$  here also supports an intercept width as $\frac{2}{N_\text{BS}}$, which is consistent with unit bandwidth in orthogonal far-field discrete Fourier transform (DFT) codebooks \cite{DFTcodebook}. Therefore, we coarsely approximate the bandwidth as $B_{k_1}\approx |k_0-k_1|N_\text{BS}+\frac{2}{N_\text{BS}}$ for consistency. According to Theorem \ref{Theorem1}, the total power at each column $k=k_1$ is fixed as $N_\text{BS}$ and corresponding interval length is $B_{k_1}$. Assume the power is uniformly distributed in the interval and thus the average coherence at each inside codeword point $(b_q,k_1)$ can be yielded to $N_\text{BS}/B_{k_1}$, i.e.,
\begin{equation}
	\arraycolsep=1.0pt\def\arraystretch{1.5}
\begin{array}{ccl}
\displaystyle g^\text{ideal}_{b_q,k_1}&=& \displaystyle\sqrt{ \frac{N_\text{BS}}{B_{k_1}}}\cdot \text{rect}\left(\frac{b_q-b_0}{B_{k_1}}\right)\\
&=&\displaystyle \left\{ \begin{array}{cl}
\displaystyle \frac{1}{\sqrt{|k_0-k_1|+2/N_\text{BS}^2}}\  &,\displaystyle \ |b_q-b_0|\leq\frac{B_{k_1}}{2}\\
0\ &, \ \text{else}
\end{array}
\right.
\end{array}
\label{ideal_pattern_gain}
\end{equation}

From above analysis we can get that, the ideal spatial-chirp beam's bandwidth inside $k=k_1$ is proportional to the distance ($|k_0-k_1|$), while the ideal average gain at each point inside $k=k_1$ is approximately inversely proportional to the distance $|k_0-k_1|$. Then we can depict this property into $k$-$b$ domain as shown in Fig. \ref{spatial_chirp_beam_pattern}, where the darkness represents coherences between $\mathbf{f}_\text{eg}(k_0,b_0)$ and the corresponding local point. The specific coverage pattern is extremely instructive for the hierarchical codebook design in the 2-dim spatial-chirp beam training. 

\begin{remark}
 In this part, we only focus on the approximated triangular-shape distribution (i.e., the ideal beam pattern in Fig. \ref{spatial_chirp_beam_pattern} ) to design each hierarchical layer's coarse coverage and training update policy. The realistic spatial-chirp beam instead of the ideal distribution is directly utilized for XL-MIMO beam training in the next subsection. Actually, the bandwidth approximation of spatial-chirp beam is not accurate when with small spatial-angular product $|k_0-k_1|D^2< \infty$. And then the realistic near-field beam will contain a roughly divergent power distribution as shown in Fig. \ref{spatial_chirp_beam_pattern}, which causes inter-beam overlapping and searching error probability. For this reason principally, in Section V we further propose an enhanced codeword generation scheme for spatial-chirp beam modification and overlapping elimination.
\end{remark}

\subsection{Spatial-chirp-based Hierarchical Beam Training}

\subsubsection{Top layer hierarchical searching}
\ \ 

\begin{figure}[!t]
\centering
\includegraphics[width=0.8\linewidth]{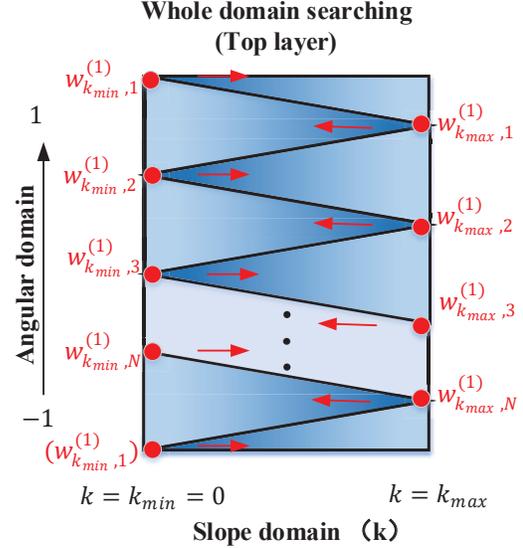}
\caption{Top-layer beam training codebook design for XL-MIMO.}
\label{top_layer_codebook}
\end{figure}

In the top-layer beam training, suppose the spatial-chirp codewords are all selected at margin of $k-b$ domain $[k_\text{min},k_\text{max}]\times [-1,1]$. When the ideal beam pattern (\ref{ideal_pattern_width}) (\ref{ideal_pattern_gain}) is assumed, to fully cover the whole $k$-$b$ domain, the corresponding spatial-chirp beam distribution should be designed as shown in Fig. \ref{top_layer_codebook}. Each spatial-chirp beam can support beam searching inside a triangular region. In Fig. \ref{top_layer_codebook} the darkness denotes beamforming gain for the corresponding channel path with coordinate at each point. Define the $i$-th codeword at column $k$ as $\bm{w}^{(\ell)}_{k,i}$, where index indicator $\ell=1$ represents the top hierarchical layer. The angular sampling spacing and codebook size at top layer should follow the next theorem:
\begin{theorem}
In top-layer beam training, angular sampling spacing is at most $B_\text{max}=\sqrt{2/N_\text{BS}}$ and thus the codebook size is at least $2\sqrt{2N_\text{BS}}$.
\label{Theorem2}
\end{theorem}
\begin{IEEEproof}
According to the near-field Fresnel region analysis \cite{minimum_distance05}, the distance lower bound of Fresnel region is $r_\text{min}=0.5\sqrt{\frac{D^3}{\lambda}}$, which is defined as the boundary value to distinguish between reactive near field and radiating near field. The realistic transmission distance should follow $r_0>r_\text{min}$, and correspondingly we have $k_\text{max}\leq \frac{\lambda}{4r_\text{min}}$. Substituting it into (\ref{ideal_pattern_width}) we  can get that 
\begin{equation}
B_\text{max}= k_\text{max}N_\text{BS}\leq \frac{\lambda}{4r_\text{min}}N_\text{BS}=\sqrt{\frac{2}{N_\text{BS}}}.
\label{top_resolution}
\end{equation}
In the top-layer searching, the codewords are all configured at two columns ($k=k_\text{min}=0$ and $k=k_\text{max}$). Inside each column we have to quantiza intercept $[-1,1]$ with intra-column spacing $B_\text{max}$, therefore we can obtain the total codebook size at top hierarchical layer as 
\begin{equation}
N^{(1)} = 2\times \frac{1-(-1)}{B_\text{max}}=2\sqrt{2N_\text{BS}}.
\label{top_word_num}
\end{equation}
\end{IEEEproof}

From Theorem \ref{Theorem2} we obtain the top-layer codebook size and angular spacing as $N^{(1)}=2\sqrt{2N_\text{BS}}$ and $B_\text{max}$. Notice that the top-layer codewords together should fully cover the whole region $[k_\text{min},k_\text{max}]\times[-1,1]$ while all codewords lie in the boundary. Therefore, at least two columns of codewords with slopes $k=k_\text{min}$ and $k=k_\text{max}$, are indispensable  in top-layer training, where each column contains $N^{(1)}/2$ codewords. Due to the triangular-shape dominant region for each codeword, the codeword groups between two columns $k=k_\text{min}$ and $k=k_\text{max}$ should keep interlaced with fixed offset spacing $B_\text{max}/2$. The detailed codeword distribution is shown in Fig. \ref{top_layer_codebook}, which can be further formulated with the coordinates of top-layer codewords as follows:
\begin{equation}
\arraycolsep=1.0pt\def\arraystretch{1.0}
\left\{	\begin{array}{llc}
\bm{w}^{(1)}_{k_\text{min},i}:&& \displaystyle \left(k_\text{min},iB_\text{max}\right); \\
\bm{w}^{(1)}_{k_\text{max},i}:&&\displaystyle  \left(k_\text{max},(i+0.5)B_\text{max}\right); 
\end{array}
\right., i=1,2,\dots, {N^{(1)}/2}
\label{top_layer_codeword}
\end{equation}
This conclusion is quite succinct, which shows that the top-layer codebook size and supporting region only depend on the number of BS antenna $N_\text{BS}$. For example, when we set carrier frequency $f_c=30\text{GHz}$ and $N_\text{BS}=512$, the minimum communication distance $r_\text{min}$ can be easily calculated as $8\text{m}$. Thus the codebook size is obtained as $N^{(1)}=64$, which seems acceptable for realistic beam training. 

Besides, it is worth noting that wider sector searching can also be similarly designed to further reduce training overhead. Instead of triangular-shape design at the top-layer here, the conventional far-field hierarchical training such as binary searching and multi-finger beam training, can also be deployed here to coarsely determine the user location for overhead reduction.

\subsubsection{Hierarchical Codeword Update Policy}
\ \ 

\begin{figure*}[!t]
	\begin{center}
		\subfigure[Conventional far-field (angular-domain) hierarchical beam training]{
			\includegraphics[width=0.6\linewidth]{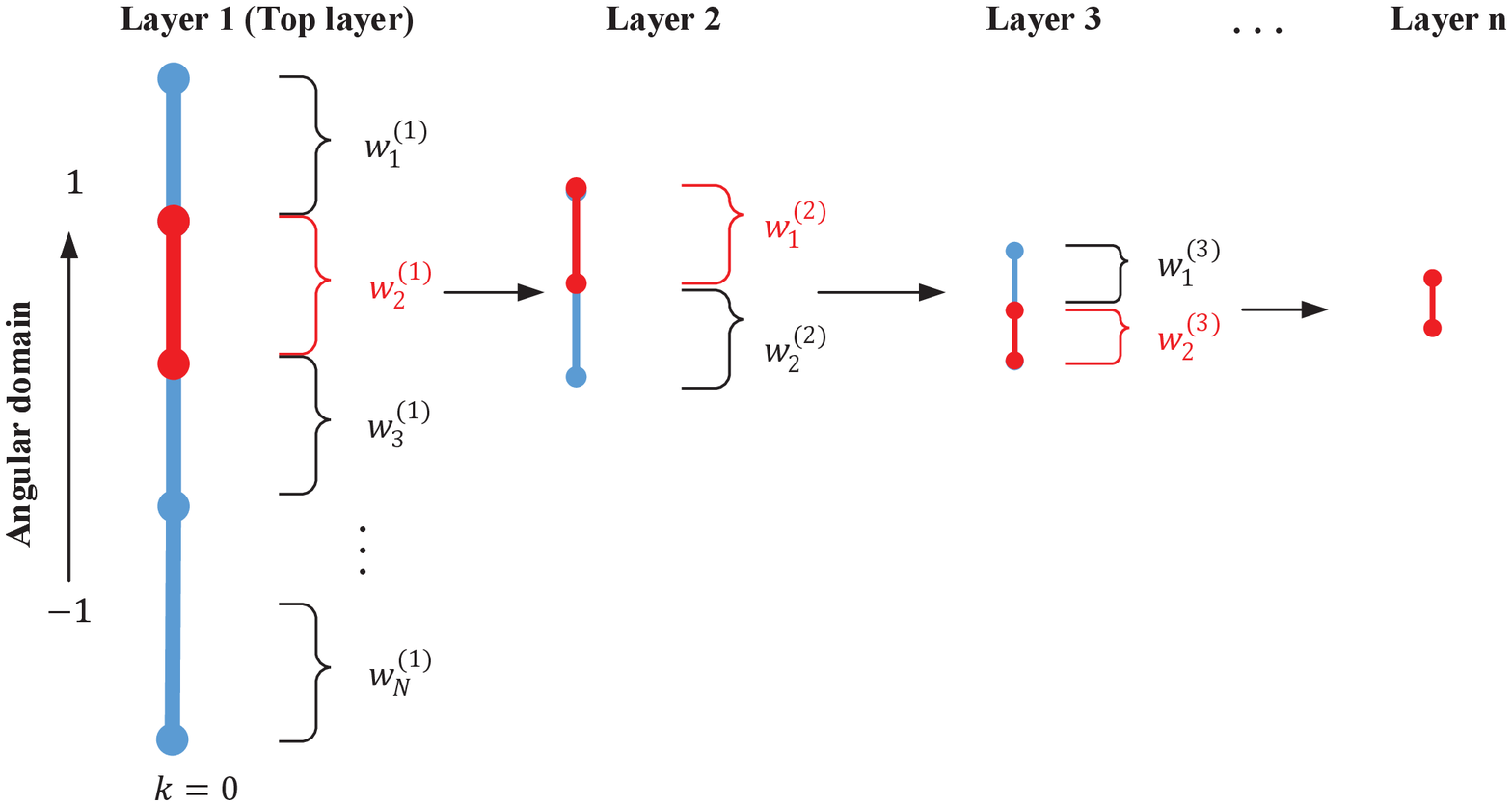}
			\label{far_field_beam_training}
		}
		\subfigure[Proposed near-filed ($k$-$b$ domain) XL-MIMO hierarchical beam training]{
			\includegraphics[width=0.9\linewidth]{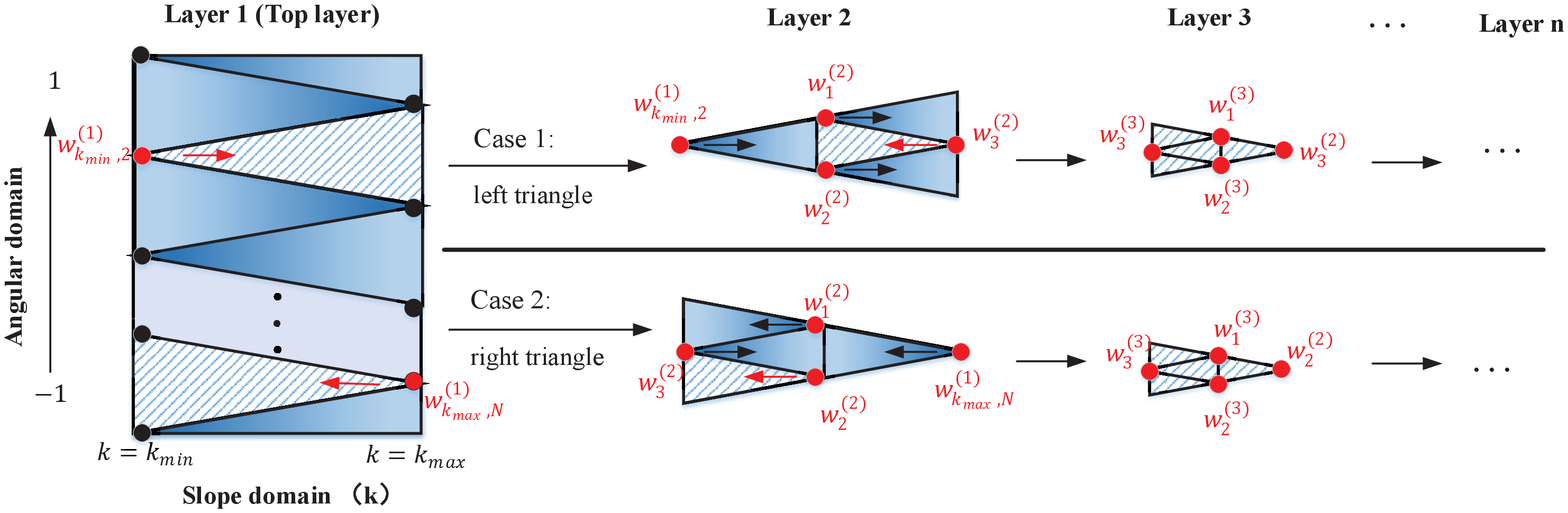}
			\label{near_field_beam_training}
		}
		\caption{The hierarchical beam training and codeword update rule for (a) conventional  far-field (angular-domain) system model; (b) near-field  ($k$-$b$ domain) XL-MIMO system.}
	\end{center}
\end{figure*}

When the top-layer beam searching is addressed, we obtain the codeword with strongest beam gain which is herein marked as $\bm{w}^{(1)}_\text{opt}$. The corresponding $k$-$b$ domain triangle region is determined as $\mathcal{R}^{(1)}_\text{opt}$. Then we should allocate additional next-layer codewords for more accurate beam training and further reduce the potential beam region $\mathcal{R}_\text{opt}$ until the region area satisfies highest resolution or approaches required beamforming gain.  

As for each selected triangle region $\mathcal{R}^{(1)}_\text{opt}$, we divide it into four specific fractions, as shown in Layer 2 in Fig. \ref{near_field_beam_training}. We'd like to discuss the problem in two cases. For Case 1 as shown in Fig. \ref{near_field_beam_training}, the top-layer optimal region is a left triangle. Without loss of generality, we assume the optimal top-layer codeword is $\bm{w}_\text{opt}^{(1)}=\bm{w}^{(1)}_{k_\text{min},2}$. Next in Layer 2, three additional codewords, i.e., $\bm{w}_1^{(2)}$, $\bm{w}_2^{(2)}$ and $\bm{w}_3^{(2)}$ are searched. The corresponding codewords' coordinates in $k$-$b$ domain can be easily obtained as middle points of the three sides of  triangle  $\mathcal{R}^{(1)}_\text{opt}$, i.e.,
\begin{equation}
\arraycolsep=1.0pt\def\arraystretch{1.0}
\left\{	\begin{array}{llc}
\displaystyle \bm{w}^{(2)}_{1}:&&\displaystyle \left(\frac{k_\text{min}+k_\text{max}}{2},(i-0.25)B_\text{max}\right); \\
\displaystyle \bm{w}^{(2)}_{2}:&&\displaystyle  \left(\frac{k_\text{min}+k_\text{max}}{2},(i+0.25)B_\text{max}\right); \\
\displaystyle \bm{w}^{(2)}_{3}:&&\displaystyle  \left(k_\text{max},\ \ iB_\text{max}\right); 
\end{array}
\right.
\label{bottom_codeword}
\end{equation}
When in this example with $\bm{w}_\text{opt}^{(1)}=\bm{w}^{(1)}_{k_\text{min},2}$, set $i=2$ in (\ref{bottom_codeword}) and the three codewords are then obtained. Notice that the temporary codeword subset $\{\bm{w}^{(1)}_{k_\text{min},2},\bm{w}_1^{(2)},\bm{w}_2^{(2)},\bm{w}_3^{(2)}\}$ can uniformly divide region $\mathcal{R}^{(1)}_\text{opt}$ into four non-overlapping sub-triangles and each codeword dominates in one sub-region, which means by comparing the four codewords' beamforming gains, we can further reduce the potential region by four times. Both $k$ dimension and $b$ dimension can be reduced by half. By several 2-dim binary searches, we can successively obtain $\mathcal{R}^{(3)}_\text{opt}$, $\dots$, $\mathcal{R}^{(l)}_\text{opt}$, $\dots$. until $l$ approaches the maximum hierarchical layer $L$ and then get the final beam training result $\bm{w}^{(L)}_\text{opt}$. For the other case (Case 2: right triangle) in Fig. \ref{near_field_beam_training}, we omit its detailed description for brevity since its process is quite similar to Case 1. The corresponding updated codewords' coordinates of Case 2 can be formulated as:
\begin{equation}
\arraycolsep=1.0pt\def\arraystretch{1.0}
\left\{	\begin{array}{llc}
\displaystyle \bm{w}^{(2)}_{1}:&&\displaystyle \left(\frac{k_\text{min}+k_\text{max}}{2},(i-0.25)B_\text{max}\right); \\
\displaystyle \bm{w}^{(2)}_{2}:&&\displaystyle  \left(\frac{k_\text{min}+k_\text{max}}{2},(i+0.25)B_\text{max}\right); \\
\displaystyle \bm{w}^{(2)}_{3}:&&\displaystyle  \left(k_\text{min},\ \ iB_\text{min}\right); 
\end{array}
\right.
\label{bottom_codeword_Case2}
\end{equation}

In retrospect, the conventional far-field beam training can be regarded as a particular case of spatial-chirp-based XL-MIMO training. As shown in Fig. \ref{far_field_beam_training}, if we only focus on the overall angular values with fixed slope $k=0$, the 2-dim hierarchical searching procedure proposed above will degrade to angular-domain 1-dim hierarchical beam training, where only two codewords are searched in each layer. From this point of view, our proposed near-field XL-MIMO hierarchical training contains only limited pilot overhead which is realistic and quite comparable to conventional mmWave binary hierarchical beam searching. 

\subsubsection{Quantization order calculation}
\ \ 

The next question is how to determine the value of maximum hierarchical layer $L$. Too large $L$ is unnecessary due to  redundant training overhead, while little $L$ may cause training error and beamforming degradation. An intuitive idea is to first configure the angular quantization resolution as $2/N_\text{BS}$ (i.e. $N_\text{BS}$ orthogonal DFT beam codewords for the whole angular domain, which is widely utilized). Since 2-dim binary search simultaneously processes on both intercept ($b$) domain and slope ($k$) domain, the quantization of slope $k$ is also determined.  Therefore it is sufficient if we can prove that the above quantization of $k$ is supportive and fully covers the whole slope interval $[0,k_\text{max}]$.

From above analysis, the maximum angular bandwidth for top-layer codeword is $B_\text{max}=k_\text{max}N_\text{BS}$ while the slope-domain interval length is $k_\text{max}$. When highest resolution of angular domain is set as $B_\text{min}=2/N_\text{BS}$, each top-layer codeword contains 
\begin{equation}
L_{b}^{(1)} = B_\text{max}\big/B_\text{min}=\frac{1}{2}k_\text{max}N_\text{BS}^2
\label{triangle_angular_num}
\end{equation}
basic quantized angular fractions and correspondingly the number of slope-domain fraction is the same $L_k^{(1)}=L_b^{(1)}$. Therefore the basic slope-domain quantized fraction length can be calculated as 
\begin{equation}
\Delta k = k_\text{max}/ L_k^{(1)} = \frac{2}{N_\text{BS}^2}.
\label{spacing_k_here}
\end{equation}

According to the property of Fresnel integral and derivation in \cite[Lemma 1]{CE_2_polardomain}, when the reciprocal of distance $1/r_0$ is uniformly quantized with quantization spacing smaller than threshold $\frac{2\lambda \beta^2}{N_\text{BS}^2d^2(1-\theta^2)}$, the beam can take a full coverage along slope domain. Substituting the threshold of quantized $1/r_0$ back to $k=\frac{\lambda (1-\theta^2)}{4r_0}$ we can directly get the corresponding slope-domain quantization spacing as 
\begin{equation}
k_\text{TH}=\Delta k^\text{upper} =  \frac{2\lambda \beta^2}{N_\text{BS}^2d^2(1-\theta^2)}\cdot \frac{\lambda(1-\theta^2)}{4}=\frac{2\beta^2}{N_\text{BS}^2}
\label{spacing_k_cmy}
\end{equation}
Compare the two spacing values $\Delta k$ (\ref{spacing_k_here}) and $\Delta k^\text{upper}$ (\ref{spacing_k_cmy}), and we can observe that the spatial-chirp-based slope quantization spacing can be designed from (\ref{spacing_k_cmy}) with $\beta=1$, and thus we have $\Delta k< \Delta k^\text{upper}$ since  the upper-bounded $\beta$ is approximately set as $1.6$ in \cite{CE_2_polardomain}. Therefore, we can confirm that $\Delta k$ here is supportive for the whole slope interval $[0,k_\text{max}]$. 

To summarize, the whole hierarchical layer $L$ can be calculated as 
\begin{equation}
L = \lceil \log_2 L_k^{(1)} \rceil=\bigg\lceil \log_2 \frac{1}{2}k_\text{max}N_\text{BS}^2 \bigg\rceil.
\label{layer_num}
\end{equation}
Besides, when $L$ is determined we can derive the top-layer codebook size as 
\begin{equation}
N^{(1)}=2\times 2^{(\log_2 N_\text{BS}-L)}=2^{1-L}N_\text{BS}.
\label{top_word_num2}
\end{equation}
Notice that this derivation (\ref{top_word_num2}) is consistent to (\ref{top_word_num}) when we directly assume the minimum distance $r_0=r_\text{min}=0.5\sqrt{\frac{D^3}{\lambda}}$ (we have $k_\text{max}=\frac{\lambda}{4r_0}=\sqrt{\frac{2}{N_\text{BS}^3}}$). And it is worth noting that the tradeoff exists when taking joint consideration of (\ref{layer_num}) and (\ref{top_word_num2}). If the minimum distance $r_0$ increases caused by environment variance, $k_\text{max}$ decreases and correspondingly, we need fewer hierarchical layers  $L$ but more codewords in top layer. The concrete steps of the proposed spatial-chirp-based hierarchical beam training scheme are displayed in Algorithm \ref{alg1} as follows.

\begin{algorithm}[htb] 
\normalem
\caption{Proposed Chirp-based hierarchical beam training for near-field XL-MIMO} 
\label{alg1} 
\begin{algorithmic}[1] 
\REQUIRE System configurations $N_\text{BS}$, $f_c$, $d$ and minimum communication distance $d_\text{min}$, required beamforming gain threshold $g_\text{th}$.

\ENSURE Beam training result $\mathbf{b}$


\STATE Calculate $k_\text{max}=\frac{\lambda}{4r_\text{min}}$, hierarchical layer $L$ (\ref{layer_num}), top-layer codebook size $N^{(1)}$ (\ref{top_word_num2}) and angular-domain spacing $B_\text{max}=k_\text{max}N_\text{BS}$

\emph{\%\% Top-layer search}
\STATE Generate top-layer codewords via (\ref{top_layer_codeword}) and search for the codeword with maximum beamforming gain $\mathbf{w}^{(1)}_\text{max}$. Mark the corresponding $k$-$b$ domain region as $\mathcal{R}^{(1)}_\text{opt}$. 

\emph{\%\% Hierarchical search}
\FOR{layer index $l=2$ to $L$}
\STATE Divide region $\mathcal{R}^{(i)}_\text{opt}$ into four fractions and generate corresponding four dominant codewords in $i$-th layer $\{\bm{w}^{(i-1)}_\text{max},\bm{w}^{(i)}_1, \bm{w}^{(i)}_2, \bm{w}^{(i)}_3\}$ via (\ref{bottom_codeword}) and (\ref{bottom_codeword_Case2})
\STATE Select the best codeword with maximum gain $\bm{w}^{(i)}_\text{max}$ and update the optimal region $\mathcal{R}^{i}_\text{opt}$
\ENDFOR
\STATE Final optimal supporting beam $\mathbf{b}\leftarrow \bm{w}^{(L)}_\text{opt}$
\end{algorithmic}
\end{algorithm}

\section{Enhancement of Chirp-based Hierarchical Codebook}

\subsection{Spatial-chirp-based Codeword Modification}

In this subsection, we propose an enhanced hierarchical codebook design method for XL-MIMO system. Based on Principle of Stationary Phase (PSP) \cite{PSP}, the numerical solution of spatial-chirp signal's angular amplitude  at $\theta=\theta_0$ can be approximately calculated as 
\begin{equation}
G(\theta_0,{\hat{\bm a}}^\text{n-f}_{k,b}) = \frac{1}{\sqrt{2k}}\sqrt{[C(X_1)+C(X_2)]^2+[S(X_1)+S(X_2)]^2}
\label{POSP_spatial}
\end{equation}
where $C(X)=\int_0^X\cos(\frac{\pi x^2}{2})dx$ and $S(X)=\int_0^X\sin(\frac{\pi x^2}{2})dx$ are Fresnel integral, and the variables $X_1$ and $X_2$ are written as follows. The results here are simple through PSP and thus we omit detailed derivation for brevity.
\begin{equation}
X_1 = \frac{kN_\text{BS}+(\theta_0-b)}{\sqrt{2k}},\ \ X_2 = \frac{kN_\text{BS}-(\theta_0-b)}{\sqrt{2k}}.
\end{equation}
Actually as shown in (\ref{POSP_spatial}), the realistic spatial-chirp beam could never keep standard triangle region like (\ref{ideal_pattern_width}) and (\ref{ideal_pattern_gain}), and the bias is vividly compared in Fig. \ref{spatial_chirp_beam_pattern}. Therefore, the overlapping among several spatial-chirp beam vectors may cause training error and thus degrade the following beamforming gain, which is still a challenging problem in the proposed spatial-chirp-based hierarchical training above. 

To approach the ideal beam pattern  (\ref{ideal_pattern_width}) and (\ref{ideal_pattern_gain}) as much as possible, we establish a general optimization framework for each hierarchical layer, with initial input as spatial-chirp beam vectors. Notice that inside each layer we only need to design one codeword $\bm{w}^{(l)}_\text{enh}$ and the rest can be directly generated by shifting by $(p_n\Delta k,q_n\Delta b)$, so we only consider one fixed initial input spatial-chirp beam $\bm{w}_0: \ (k,b)=(0,0)$. 

For the $l$-th layer, the total ideal dominant region $\mathcal{\hat{R}}^{(l)}_\text{ideal}$ contains two sub-regions, i.e., the combination of right triangle and left triangle. Since the dominant direction (right or left) is controlled by the $(l-1)$-th layer's searching result, the two cases may both appear and thus we have to take joint consideration of the two directions. The corresponding triangle side ($b$-axis) $B_\text{max}^{(l)}$ and height ($k$-axis) $k^{(l)}$ are yielded to
\begin{equation}
B^{(l)}_\text{max} = k_\text{max}N_\text{BS}2^{1-l}, \ \ k^{(l)}=k_\text{max}2^{1-l}
\label{side_height}
\end{equation}
Following the spatial-chirp-based quantization order and codeword distribution in Algorithm 1,  we can discretize the ideal region $\mathcal{\hat{R}}^{(l)}_\text{ideal}$ into several bottom-layer codewords.
Inside  $l$-th layer region $\mathcal{\hat{R}}^{(l)}_\text{ideal}$ there exist $2L_k^{(l)}-1$ ($L_k^{(l)}=2^{L-l+1}$ ) columns $k_p^{(l)}=p\Delta k,p =1-L_k^{(l)},\dots,-1,0,1,\dots,L_k^{(l)}-1$, and corresponding codebook size is $L_k^{(l)},\dots,2,1,2,\dots,L_k^{(l)}$, respectively. Therefore the ideal gain (coherence with quantized beam codeword $(k_p^{(l)},b_q)$) for layer $l$ is formulated as
\begin{equation}
	\arraycolsep=1.0pt\def\arraystretch{1.5}
\begin{array}{ccl}
&&\displaystyle G_l(k_p^{(l)},b_q) =\left\{ \begin{array}{cl}
\displaystyle \sqrt{\frac{N_\text{BS}}{|p|+1}}\  &,\displaystyle \ |b_q|\leq N_\text{BS}|k_p^{(l)}|\\
0\ &, \ \text{else}
\end{array}
\right.,\\
&&\ \ \ \ \ \ \ \ \ \ \ k_p^{(l)}=p\Delta k,\ p =1-L_k^{(l)},\dots,L_k^{(l)}-1
\end{array}
\label{ideal_G_l}
\end{equation}

Inside the $l$-th layer, the optimization objective is established as follows to reduce the gap between realistic beam vector and ideal beam pattern:
\begin{equation}
\arraycolsep=1.0pt\def\arraystretch{1.5}
\begin{array}{ccc}
\displaystyle \min_{\mathbf{b}^{(l)}} &&\displaystyle\ \ \  \left\|\ \ \left(\mathbf{r}_l-|\mathbf{A}_l^H\mathbf{b}^{(l)}| \right)\circ \bm\phi_l\ \ \right\|^2\\
\text{s.t.}&& |\mathbf{b}^{(l)}(n)|=1,\ \forall \ n=1,\dots,N_\text{BS}
\end{array}
\label{init_objective}
\end{equation}
where  $\mathbf{A}_l=\text{mat}(\mathcal{A}_l), \ \mathcal{A}_l\in \mathbb{C}^{N_\text{BS}\times N_\text{BS}\times (2L_k^{(l)}-1)}$ is the beam steering tensor with overall quantized beam vectors inside region $[-k^{(l)},k^{(l)}]\times [-1,1]$ into one group, and $\mathbf{r}_l=\text{vec}(\mathbf{G}_l), \mathbf{G}_l=[G_l(k_q^{(l)},b_q)]\in \mathbb{C}^{N_\text{BS}\times (2L_k^{(l)}-1)}$ denotes the perfectly ideal quantized coherence vector in this region. Besides, $\bm\phi_l=\text{vec}(\bm\Phi_l),\bm\Phi_l\in \mathbb{R}^{N_\text{BS}\times(2L_k^{(l)}-1)}$ is a fixed weighting matrix for $l$-th layer's codeword design.  Generally, we can set $\bm\Phi_l=\bm 1_{N_\text{BS}\times(2L_k^{(l)}-1)}$ for equitable weights. On the other hand, notice that  the overlapping among spatial-chirp-based codewords usually appears at two sides $k=\pm k^{(l)}$ as shown in Fig. \ref{spatial_chirp_beam_pattern}. We tend to design the weighting matrix $\bm \Phi_l$ with larger weights at two sides as follows
\begin{equation}
\bm\Phi_l =\underbrace{ [L_k^{(l)},\dots 3,2,1,2,3,\dots,L_k^{(l)}] }_{2L_k^{(l)}-1 \ \ \text{values}}\otimes \bm 1_{N_\text{BS}\times 1}.
\end{equation}
By introducing an auxiliary phase vector $\bm\psi^{(l)}$ the previous objective (\ref{init_objective}) is reformulated as 
\begin{equation}
\arraycolsep=1.0pt\def\arraystretch{1.5}
\begin{array}{ccc}
\displaystyle \min_{\mathbf{b}^{(l)},\bm\psi^{(l)}} &&\displaystyle\ \ \  \left\|\ \ \left(\mathbf{r}_l\circ e^{j\bm\psi^{(l)}}-\mathbf{A}_l^H\mathbf{b}^{(l)} \right)\circ \bm\phi_l\ \ \right\|^2\\
\text{s.t.}&& |\mathbf{b}^{(l)}(n)|=1,\ \forall \ n=1,\dots,N_\text{BS}
\end{array}
\label{objective_2}
\end{equation}
and then the problem can be easily solved by alternating minimization. With $\mathbf{b}^{(l)}$ fixed we can easily obtain the closed-form solution of $\bm\psi^{(l)}$. When we fix $\bm\psi^{(l)}$ to optimize codewords, we utilize manifold gradient optimization to deal with the non-convex constant-modulus constraint. The spatial-chirp beams in Algorithm 1 are utilized here to initialize the alternating minimization scheme. 

Firstly, with codeword vector $\mathbf{b}^{(l)}_t$ fixed where $t$ denotes iteration index,  the closed-form optimal solution of auxiliary phase vector $\bm\psi_{t}^{(l)}$ is yielded to 
\begin{equation}
\bm\psi_{t}^{(l)} = \angle \mathbf{A}_l^H\mathbf{b}_{t}^{(l)} 
\label{phase_vec_update}
\end{equation}

Secondly, with phase vector $\bm\psi^{(l)}_{t}$ fixed, we project the variable $\mathbf{b}^{(l)}$ into complex unit-circle Riemannian manifold \cite{matrix_manifold} $\mathcal{M}_\text{cc}^{N_\text{BS}}=\{\mathbf{b}\in \mathbb{C}^{N_\text{BS}}: |b(1)|=\dots=|b(N_\text{BS})|=1\}$. The tangent space at one point  on the manifold $\mathbf{b}\in\mathcal{M}_\text{cc}^{N_\text{BS}}$ is formulated as 
\begin{equation}
T_{\mathbf{b}}\mathcal{M}_\text{cc}^{N_\text{BS}}=\{\mathbf{t}\in \mathbb{C}^{N_\text{BS}}: \mathbf{t}(n)\mathbf{b}(n)^*+\mathbf{t}(n)^*\mathbf{b}(n)=0\}.
\end{equation}
Then the corresponding Riemannian gradient $\text{grad}_{\mathcal{M}}f(\mathbf{b})$ is formulated as 
\begin{equation}
\text{grad}_{\mathcal{M}}f(\mathbf{b}) = \text{Proj}_{\mathbf{b}}(\nabla f(\mathbf{b}))= \nabla f(\mathbf{b}) -\Re (\nabla f(\mathbf{b}) \circ \mathbf{b}^*) \circ \mathbf{b},
\label{Riemannian_grad}
\end{equation}
and the corresponding Euclidean gradient of objective in (\ref{objective_2}) is given as 
\begin{equation}
\nabla f(\mathbf{b}^{(l)})= \mathbf{A}_l \left[\bm\phi_l^2\circ \left(\mathbf{A}_l^H\mathbf{b}^{(l)}-\mathbf{r}_l\circ e^{j\bm\psi^{(l)}_t}\right)\right]
\label{Euclidean_grad}
\end{equation}

When we calculate gradient step size $\mu_{t}$ via Armijo rule \cite{matrix_manifold}, the codeword vector can be further updated:
\begin{equation}
\mathbf{\bar{b}}^{(l)}_t = \mathbf{b}^{(l)}_t-\mu_{t} \cdot \text{grad}_{\mathcal{M}} f(\mathbf{b}^{(l)}_t).
\label{codeword_update_gradient}
\end{equation}
Notice that the temporary result $\mathbf{\bar{b}}^{(l)}_t$ doesn't satisfy the constant-modulus constraint, i.e., $\mathbf{\bar{b}}^{(l)}_t \notin \mathcal{M}_\text{cc}^{N_\text{BS}}$. \emph{Retraction} procedure is necessary to fine-tune and map the temporary result on the tangent space to the complex circle manifold. According to \cite{matrix_manifold}, the retraction operator is utilized as follows
\begin{equation}
\mathbf{b}^{(l)}_{t+1}=\text{Retr}_{\mathbf{b}^{(l)}_t}\left(\mu_{t} \cdot \text{grad}_{\mathcal{M}} f(\mathbf{b}^{(l)}_t)\right)=\text{vec}\left[ \frac{\mathbf{\bar{b}}^{(l)}_t (n)}{|\mathbf{\bar{b}}^{(l)}_t(n)|} \right].
\label{retraction_codeword}
\end{equation}
After several iterations ($t=1,\dots,T$) we obtain the near-optimal beamforming vector $\mathbf{b}_\text{opt}^{(l)}=\mathbf{b}_{T}^{(l)}$ for XL-MIMO.  Notice that in each hierarchical layer, we only consider one codeword with coordinate $(k,b)=(0,0)$ and $\mathbf{b}^{(l)}_\text{opt}$ is just the enhancement for it. Therefore in the $l$-th layer, we can modulate the enhanced basis  $\mathbf{b}^{(l)}_\text{opt}$  to overall spatial-chirp-based codewords and derive the final enhanced codebook via (\ref{bottom_codeword}), (\ref{bottom_codeword_Case2}) and  $\mathbf{b}^{(l)}_\text{opt}$  as 
\begin{equation}
\bm{w}_{\text{enh},i}^{(l)}= \bm{b}^{(l)}_\text{opt}\circ \bm{w}_i^{(l)}
\label{codeword_fine_tune}
\end{equation}
The detailed procedure above is summarized in Algorithm \ref{alg2}, which can be regarded as one enhanced codebook design for near-field XL-MIMO system. 

\begin{algorithm}[htb] 
\normalem
\caption{Proposed enhanced codebook design for XL-MIMO hierarchical beam training} 
\label{alg2} 
\begin{algorithmic}[1] 
\REQUIRE Parameters in Alg 1, maximum iteration number $T$.

\ENSURE Beam training result $\mathbf{b}$


\emph{\%\% Initialization}

\STATE Generate overall spatial-chirp-based codewords in all hierarchical layers $\{\bm{w}^{(l)}_i,i=1,2,\dots\}$, $l=1\dots,L$. Set $\mathbf{b}^{(l)}_1=\bm 0$, i.e., $(k,b)=(0,0)$ and calculate ideal beam gain distribution $\mathbf{r}_l$ and supporting matrix $\mathbf{A}_l$ in (\ref{init_objective}).

\emph{\%\% Alternating Minimization}

\FOR{hierarchical layer $l=1$ to $L$}
\FOR{iter num $t=1$ to $T$}

\STATE Update auxiliary phase vector $\bm\psi^{(l)}_t$ via (\ref{phase_vec_update}) 

\STATE Update codeword vector $\mathbf{b}^{(l)}_{t+1}$ through manifold gradient optimization via (\ref{Riemannian_grad}), (\ref{Euclidean_grad}), (\ref{codeword_update_gradient}),  (\ref{retraction_codeword})

\ENDFOR

\STATE Calculate the enhanced codebook $\{\bm{w}^{(l)}_{\text{enh},i},i=1,2,\dots\}$ in the $l$-th hierarchical layer via (\ref{codeword_fine_tune})

\ENDFOR

\emph{\%\% Hierarchical beam training for XL-MIMO}

\STATE Utilize the enhanced codebook for hierarchical beam training following Algorithm 1 (Step 3, 4 and 5).

\STATE Final optimal supporting beam $\mathbf{b}\leftarrow \bm{w}^{(L)}_{\text{enh},\text{opt}}$
\end{algorithmic}
\end{algorithm}

\subsection{Fine-tune for the Realistic Communication}
Limited by the realistic communication environment and BS equipment, there usually exists one minimum communication distance $r_0\geq r_\text{min}\geq 0.5\sqrt{\frac{D^3}{\lambda}}$ as above analysis, which can be defined as the boundary value to distinguish between reactive near field and radiating near field.  Notice that in Section III we assume that $k$ and $b$ are mutually independent to avoid additional  complexity. Actually the slope is partially controlled by intercept. For example, when the AoD is fixed as $\theta_0=\pm 1$, the transmission distance for each antenna element is $r_n=r_0+nd$ and thus the steering vector is $[\mathbf{a}]_n=e^{-j\pi n}$, i.e., $k=0$ always holds and coordinate $(k,b)$ with $k\neq 0, b=\pm 1$ is impossible. Therefore we can further cut off the feasible $k$-$b$ domain region to reduce training overhead in this scenario. 

\begin{figure}[!t]
\centering
\includegraphics[width=0.6\linewidth]{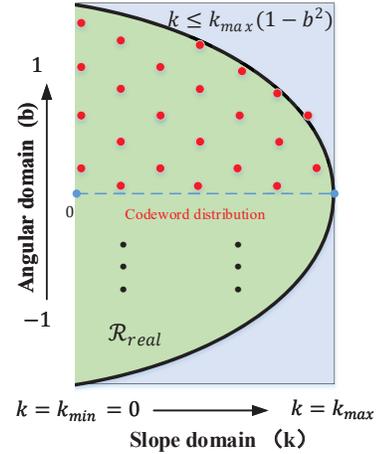}
\caption{Modified $k$-$b$ domain codeword distribution for the realistic communication, where the minimum communication distance is preset.}
\label{minimum_distance_constraint}
\end{figure}

First, given the minimum transmission distance $r_\text{min}$, we can derive and compress the accurate supportive region of $k$-$b$ domain with additional constraint (\ref{constraint_k_b}) as shown in Fig. \ref{minimum_distance_constraint}. The whole $k$-$b$ region satisfying  (\ref{constraint_k_b}) is defined as $\mathcal{R}_\text{real}$.
\begin{equation}
k\leq \frac{\lambda}{4r_\text{min}}(1-b^2)
\label{constraint_k_b}
\end{equation}

Next, for codebooks $\{\bm{w}^{(l)}_{\text{enh},i}\}$ in each hierarchical layer $l=1,\dots,L$, we delete the codewords with all dominant region outsides ($\mathcal{R}^{(l)}_i \cap \mathcal{R}_\text{real}=\varnothing$) and only retain the rest codewords which contain sub-region inside constraint (\ref{constraint_k_b}), i.e., the codewords with region intersection $\mathcal{R}^{(l)}_i \cap \mathcal{R}_\text{real}\neq \varnothing$. In this way, the codebook size can be further reduced for all hierarchical layers. Besides, the proposed hierarchical beam training schemes for XL-MIMO is also promising and applicable in near-field RIS CSI acquisition, with joint consideration of previous estimation methods such as \cite{RIS_shixu}. 
We only consider the phase change between spherical and planar wavefront in this article, and we leave the amplitude change for more accurate training method design in the future work. The detailed channel model and appliable region analysis such as the effective Rayleigh distance and equi-power line can be further found in \cite{effective_Ray,sunshu}.

\subsection{Performance Analysis}

In this subsection we provide some performance evaluations about computational complexity of codebook generation and  pilot consumption in hierarchical XL-MIMO beam training.   

As for the pilot overhead consumption, at the top-layer there exist $2\times \frac{N_\text{BS}}{L_b^{(1)}}=2^{2-L}N_\text{BS}$ codewords for traversal search. And then at each layer from $2$ to $L$, only four codewords are needed to fully cover the strongest beam in previous layer. Therefore, the overall number of training timeslots can be formulated as
\begin{equation}
P \leq  2^{2-L}N_\text{BS} + 4(L-1).
\label{pilot_length}
\end{equation}
and it is worth noting that the overhead at the top layer can be further reduced through beams with wider coverage. The corresponding beam pattern is slightly different but the codebook design and update rule are quite similar, as we described under (\ref{top_layer_codeword}). Even though we utilize  the upper bound (\ref{pilot_length}) for comparison, the excellent superiority still exists. For example, when we set carrier frequency $f_c=50\text{GHz}$, minimum communication distance $r_\text{min}=12\text{m}$ and $N_\text{BS}=512$, we have $L = 5$ and corresponding pilot consumption is $P=80$. As for conventional overall traversal search of $k$-$b$ domain, the pilot consumption will increase to $8192$. As for certain distance-based hierarchical training such as \cite{weixiuhong}, assume the maximum transmission distance $r_\text{max}=1\ \text{km}$ and distance division spacing $\Delta r=1\ \text{m}$, overall $12516$ pilots are even demanded to keep satisfying beamforming performance.

As for the computational complexity of enhanced codebook design, in each layer we only need to optimize one beamforming vector while other codewords can be directly obtained via beam rotation. The detailed complexity mainly lies in the matrix multiplications of manifold gradient optimization with $\mathcal{O}(TN_\text{BS}(2L^{(l)}_k-1))$ and thus the overall computational complexity is yielded to 
\begin{equation}
\mathcal{O}\left[\sum_{l=1}^L TN_\text{BS}^2\left(2L^{(l)}_k-1\right)\right]=\mathcal{O}\bigg[TN_\text{BS}^2\left(2^L-1-L\right)\bigg].
\end{equation}
It is worth noting that, whatever codebook is selected here, the codebook must be pre-defined at BS side and the optimization of the codebooks should also be processed before training at BS side. Therefore, it is unnecessary to focus too much on the complexity of algorithm 2 since it does not carry about any delay or hardware cost during beam training procedure. 

\section{Simulation Results}

In this section, numerical results are presented to verify the beamforming performance of our proposed spatial-chirp-based hierarchical beam training schemes. 

In the simulation we assume that the central frequency is $f_c=50\ \text{GHz}$ and the number of BS antennas is $N_\text{BS}=512$. One single-antenna user is aligned to BS via beamforming for wireless communication. The LoS path gain obeys $\beta_\text{LoS}\sim \mathcal{CN}(0,1)$ while the NLoS paths follow the i.i.d. Gaussian distribution $\beta_l\sim \mathcal{CN}(0,10^{-3}),l=1,2,3$. Based on the above configuration, we can easily calculate the minimum transmission distance $r_\text{min}=0.5\sqrt{\frac{D^3}{\lambda}}=12.29\text{m}$ and the effective Rayleigh distance $r_\text{Ray}^\text{eff}= \epsilon\sqrt{ \frac{D^2}{\lambda}}\approx 136\text{m}$, where we assume the $80\%$ beamforming gain as the effective Rayleigh distance boundary value, i.e., $\epsilon\approx 0.174$. When the distance $r_0$ is in the middle $r_\text{min}\leq r_0\leq r_\text{Ray}^\text{eff}$, the near-filed XL-MIMO channel characteristic dominates,  while when $r_0$ turns larger $r_0>r_\text{Ray}^\text{eff}$, the conventional planar-wavefront far-field channel characteristic will be dominant. Therefore, without loss of generality, we generate the transmission distance $r_0$ between BS array center and UE with uniform distribution $r_0\sim \mathcal{U}([13\text{m},150\text{m}])$. Similarly, the relative direction $\theta_0$ is uniformly generated insides interval $[-1,1]$. Then we can mathematically calculate the top-layer codebook size as $N^{(1)}=64$ while the overall number of hierarchical layers is $L=5$. 

As comparison, we herein simulate several other beam training schemes. Firstly, the beamforming scheme with perfect CSI is provided as the absolute upper bound. Besides, we collect and traverse the overall spatial-chirp-based bottom-layer codewords (exhausting searching without hierarchy) for CSI acquisition as a much tighter beamforming upper bound. Nevertheless, almost $N_\text{BS}\times 2^{L-1}=8192$ overhead length is needed in traversal which is obviously unacceptable in realistic wireless communication. And the far-field codebook is also compared where we adopt and traverse the typical angular-orthogonal DFT codebook \cite{DFTcodebook}. Since the far-field hierarchical searching obviously cannot achieve higher performance compared with traversal of DFT codebook, we neglect the simulation of conventional far-field (planar-wave) hierarchical searching such as JOINT or EJOINT. Besides, we also simulate near-field distance-based searching methods \cite{weixiuhong} where transmission distance  are uniformly divided into several fractions for XL-MIMO codebook design.

\begin{figure}[!t]
	\centering
	\includegraphics[width=0.8\linewidth]{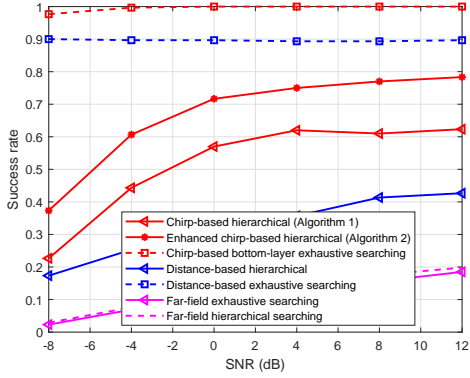}
	\caption{Success rate of spatial-chirp-based and enhanced hierarchical searching schemes for near-field XL-MIMO, with $N_\text{BS}=512$ and $f_c=50\ \text{GHz}$. Transmission distance $r_0$ and direction $\theta_0$ are uniformly generated inside $[13 \text{m},100 \text{m}]$ and $[-1,1]$, respectively.}
	\label{SuccessRate_SNR}
\end{figure}

\begin{figure}[!t]
	\centering
	\includegraphics[width=0.8\linewidth]{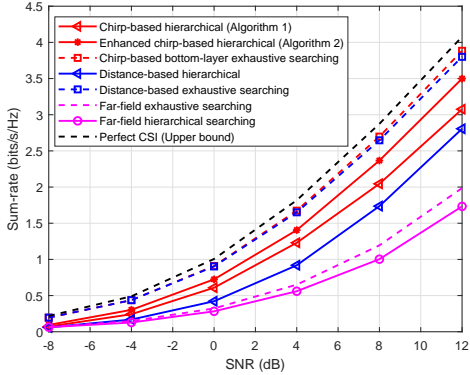}
	\caption{Comparisons of average sum-rate against SNR for several hierarchical searching schemes in near-field XL-MIMO, with $N_\text{BS}=512$ and $f_c=50\ \text{GHz}$.}
	\label{SumRate_SNR}
\end{figure}

Fig. \ref{SuccessRate_SNR} and Fig. \ref{SumRate_SNR} present the beam training success rate and communication sum-rate performances under different SNRs, respectively. The training success rate here is defined as follows. If the accurate LoS path information is captured, or the beamforming gain via hierarchical searching can approach $80\%$ under perfect CSI, we regard the hierarchical scheme works successfully. In Fig. \ref{SuccessRate_SNR} and Fig. \ref{SumRate_SNR} we can observe that the elementary codebook (overall bottom-layer codewords) works well up to $100\%$ success rate, which demonstrates the superiority of our proposed codewords. Nevertheless, the hierarchical searching couldn't obtain absolutely $100\%$ success rate since there still exists overlapping among hierarchical codewords. In fact, the ideal beam pattern $\mathbf{G}_l$ in (\ref{ideal_G_l}) is non-existent in practice and our proposed two hierarchical codebooks are just certain approximations of it. The mistakes are unavoidable during near-field hierarchical searching process. 
In detail, we can measure the overlapping via a percentage of dominant sampling points inside the ideal triangular-shape region, which can be formulated as:
 	\begin{equation}
 	\Xi(\bm w_n^{(l)})=\frac{|\mathcal{R}(w_n^{(l)})\cap \bar{\mathcal{R}}(w_n^{(l)})|}{|\mathcal{R}(w_n^{(l)})|},
 	\end{equation}
 where $\mathcal{R}(w_n^{(l)})$ denotes the realistic dominant region in (11) and $\bar{\mathcal{R}}(w_n^{(l)})$ denotes the ideal triangular-shape dominant region. $|\cdot|$ here represents the dominant region area, or the sampling point numbers in Monte Carlo simulations. Via the manifold optimization, the enhanced codebook can achieve a right dominant percentage as $95\%$, while the initial spatial-chirp-based codebook can only reach almost $85\%$. The less overlapping among hierarchical codewords world further benefits the hierarchical update success probability, and finally achieves stronger training performance and beamfocusing gain.
 Correspondingly in the simulation, the success rate can be obviously improved via enhancement (Algorithm 2) compared with the initial spatial-chirp-based scheme (Algorithm 1), and the average sum-rate is supportive and outperforms conventional far-field training by almost $1.5\ \text{bits/s/Hz}$. And our proposed hierarchical schemes outperform distance-based training method \cite{weixiuhong} by over $0.7\ \text{bits/s/Hz}$. 
 This is reasonable that distance rings in near field is nonuniform \cite{CE_2_polardomain}, which means the uniform quantization for physical spatial domain is unfair for different locations. For example, if the user locates quite near to BS, it is more sensitive to the transmission distance and thus the partition in this case should be with smaller quantization spacing. On the contrary, if the user locates quite far from BS, it only depends on the angular value but distance-insensitive, therefore, in this case the spatial quantization is unnecessary. Besides, the spatial 2-dim binary searching in \cite{weixiuhong}
cannot exactly match the three criteria above, which may further cause hierarchical update failure and training degradation.

 \begin{figure}[!t]
 	\centering
 	\includegraphics[width=0.8\linewidth]{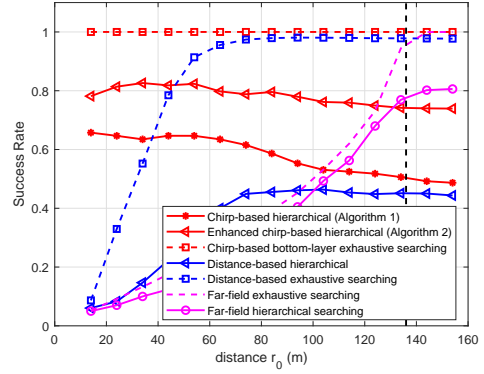}
 	\caption{Success rate of spatial-chirp-based and enhanced hierarchical searching schemes for near-field XL-MIMO, with $\text{SNR}=10\ \text{dB}$, $N_\text{BS}=512$ and $f_c=50\ \text{GHz}$. Transmission  direction $\theta_0$ is uniformly generated inside $[-1,1]$.}
 	\label{SuccessRate_distance}
 \end{figure}
 
 \begin{figure}[!t]
 	\centering
 	\includegraphics[width=0.8\linewidth]{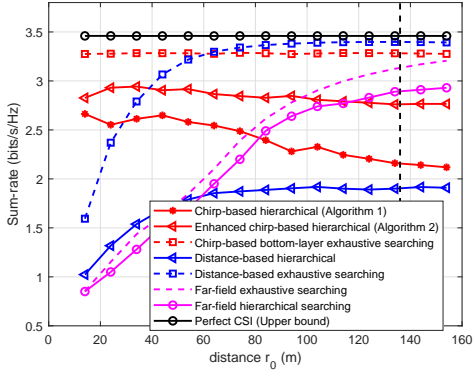}
 	\caption{Comparisons of average sum-rate against transmission distance for several hierarchical searching schemes in near-field XL-MIMO, with $\text{SNR}=10\ \text{dB}$, $N_\text{BS}=512$ and $f_c=50\ \text{GHz}$.}
 	\label{SumRate_distance}
 \end{figure}

 The impacts of transmission distance are shown in Fig. \ref{SuccessRate_distance} and Fig. \ref{SumRate_distance}, where we fix SNR as $10\ \text{dB}$. 
 We plot the effective Rayleigh distance $r_\text{Ray}^\text{eff}=136\text{m}$ with black dotted line in Fig. \ref{SuccessRate_distance} and Fig. \ref{SumRate_distance} as the boundary value between far field and spherical-wavefront near field.
 As shown here, the proposed spatial-chirp-based hierarchical training and enhanced hierarchical searching are both distance-insensitive. In another word, the proposed two hierarchical training schemes can support both far-filed and near-field scenarios with high sum-rate and beamforming gain, while the conventional DFT codebook can only provide strong support when $r_0>r_\text{Ray}^\text{eff}$. In near-field scenario such as $r_0=30\text{m}$, the DFT codebook can only approach sum-rate  $1.5 \text{bits/s/Hz}$ but the proposed schemes outperform it by almost two times. And it is worth noting that,  the enhanced spatial-chirp-based hierarchical scheme shows whole-distance superior performance which seems promising to replace conventional far-field training \cite{DFTcodebook} or distance-based hierarchical searching \cite{weixiuhong} for higher communication robustness.

\begin{figure}[!t]
	\centering
	\includegraphics[width=0.8\linewidth]{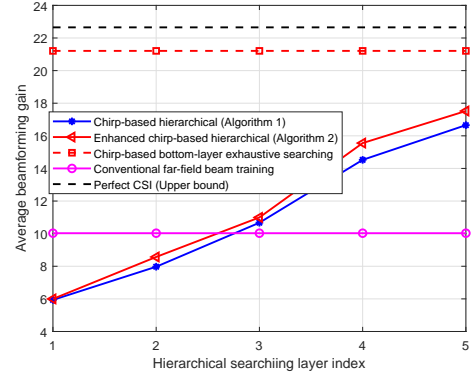}
	\caption{Average beamforming gain during each search step (each layer) for several hierarchical beam training schemes.}
	\label{BFgain_layer}
\end{figure}

Next, we provide the beamforming gain during each hierarchical layer in Fig. \ref{BFgain_layer}. At the beginning (first three upper layers), initial spatial-chirp-based scheme obtains quite comparable beamforming gain to the enhanced scheme. And then in the next two bottom layers, the gap between enhanced codebook (Alg.2) and initial spatial-chirp-based codebook (Alg.1) gets large to $6.25\%$. Compared with conventional DFT codebook, the average beamforming gain is further improved by $60\%$. Frankly speaking, from beamforming gain perspective, there still exists a large gap between the enhanced spatial-chirp-based hierarchical scheme and exhaustive searching method. This can be resolved via hybrid precoding and further improvement of update policy, which we leave in the future work.

\begin{figure}[!t]
	\centering
	\includegraphics[width=0.8\linewidth]{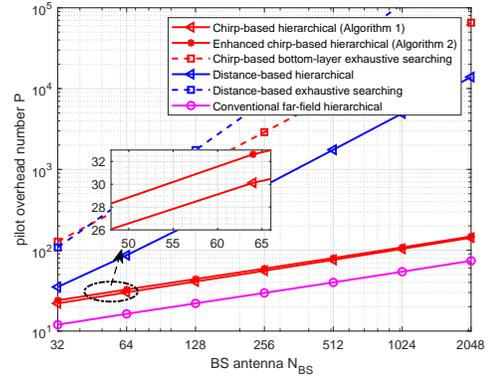}
	\caption{Overall overhead consumption against the number of BS antennas.}
	\label{overhead_BSantenna}
\end{figure}

Most importantly, we elaborate the pilot overhead consumption under different BS antenna configurations in Fig. \ref{overhead_BSantenna}. Compared with the bottom-layer overall codewords exhaustive searching, the overhead consumption can be reduced by over $99.5\%$, and over $97\%$ overhead can be saved compared with distance-based hierarchical searching method \cite{weixiuhong}. For example, when the antenna $N_\text{BS}=512$, the exhaustive searching needs overhead of about $8192$, while the enhanced spatial-chirp-based hierarchical scheme only needs $2^{2-L}N_\text{BS}+4(L-1)=80$ overhead. Besides, the initial spatial-chirp-based scheme (Algorithm 1) can further reduce the overhead consumption to $2^{2-L}N_\text{BS}+3(L-1)=76$, this is because in all hierarchical layers it utilizes the same standard spatial-chirp beams with different coordinates and the optimal codeword $\mathbf{w}^{(l)}_\text{opt}$ in layer $l$ is reused in the next $(l+1)$-th layer. Besides, it is worth noting that the main overhead consumption is from the top-layer searching where $64$ overhead is allocated here and it can be further reduced through beams with wider trapezoidal coverage above top-layer searching. 
If the far-field hierarchical searching starts with the same angular resolution $B_\text{max}$ (\ref{top_resolution}), the total overhead can be calculated as $2^{1-L}N_\text{BS}+2(L-1)=40$. Therefore, we can conclude that compared with conventional DFT codebook and corresponding hierarchical schemes, our proposed near-field hierarchical searching for XL-MIMO can approach much more superior and robust performance with quite comparable overhead consumption.

\section{Conclusion}

In this paper, we propose two low-overhead hierarchical beam training schemes for near-field XL-MIMO system to rapidly capture CSI with low training overhead consumption. Firstly, spatial-angular domain representation and slope-intercept domain representation are proposed to describe near-field channel. Based on the slope-intercept domain representation, we point out three critical criteria for XL-MIMO hierarchical beam training. Then inspired by the LFM Radar waveform, a novel spatial-chirp beam-aided codebook and corresponding hierarchical update policy are proposed. Furthermore, given the imperfect coverage and overlapping of spatial-chirp beams, we further design an enhanced hierarchical training codebook via manifold optimization and alternative minimization. Thankfully, the simulation confirms the remarkable beam training performance with reduced overhead for the proposed scheme.


%

%
%
%
%
%

\ifCLASSOPTIONcaptionsoff
  \newpage
\fi

\bibliographystyle{ieeetr}
\normalem
\bibliography{bare_jrnl.bib}

\vspace{12pt}

\end{document}